\documentclass[amsmath,amssymb,aps,showkeys,floatfix,prd,a4paper,
superscriptaddress]{revtex4}

\usepackage{dcolumn}
\usepackage{bm}
\usepackage{epsfig}
\usepackage{amsfonts}
\usepackage{amssymb,amscd}
\usepackage{multirow}

\usepackage{xcolor}
\usepackage[normalem]{ulem} 

\usepackage{tikz}
\usepackage{tikz-feynman}
\tikzfeynmanset{compat=1.1.0}
\tikzfeynmanset{warn luatex=false}
 
\newcommand{\be}{\begin{equation}}
\newcommand{\ee}{\end{equation}}
\def\beq{\begin{equation}}
\def\eeq{\end{equation}}
\def\beqa{\begin{eqnarray}}
\def\eeqa{\end{eqnarray}}
\newcommand{\ba}{\begin{eqnarray}}

\begin{document}

\title{Probing the $X(3700)$ through two-photon $D\bar{D}$ production in 
ultraperipheral collisions and at the EIC}

\author{F.C. Sobrinho}
\affiliation{Instituto de F\'\i{}sica, Universidade de S\~ao Paulo,
Rua do Mat\~ao 1371 - CEP 05508-090,
Cidade Universit\'aria, S\~ao Paulo, SP, Brazil}

\author{F.S. Navarra}
\affiliation{Instituto de F\'\i{}sica, Universidade de S\~ao Paulo,
Rua do Mat\~ao 1371 - CEP 05508-090,
Cidade Universit\'aria, S\~ao Paulo, SP, Brazil}

\author{K.P. Khemchandani}
\affiliation{Universidade Federal de S\~ao Paulo,
CEP 01302-907, S\~ao Paulo, SP, Brazil}

\author{A. Mart\'\i{}nez Torres}
\affiliation{Instituto de F\'\i{}sica, Universidade de S\~ao Paulo,
Rua do Mat\~ao 1371 - CEP 05508-090,
Cidade Universit\'aria, S\~ao Paulo, SP, Brazil}

\begin{abstract}
We investigate whether photon-induced production in ultraperipheral heavy-ion collisions and at the EIC can provide a new environment to search for the scalar $D\bar D$ molecular candidate usually denoted as $X(3700)$. We calculate the tree-level amplitudes for $\gamma\gamma\to D^+D^-$ and $\gamma\gamma\to D^0\bar D^0$, dress the outgoing pair with a coupled-channel Bethe-Salpeter interaction in the $\{D^+D^-,D^0\bar D^0\}$ basis, and convolute the resulting subprocess cross sections with equivalent-photon spectra. In the adopted two-channel scenario the analytic continuation of the $T$-matrix gives a bound state at $M_X=3.7162$~GeV, which is below the $D^0\bar D^0$ threshold. As a consequence of this, the observed signal in the continuum is not simply an isolated narrow peak, and
a characteristic near-threshold line-shape distortion is produced: around $3.70\leq W\leq3.85$~GeV the integrated dressed-to-bare ratios are $(13.71,0.35)$ in Pb-Pb UPCs and $(13.74,0.35)$ in $e$-Au collisions for $(D^+D^-,D^0\bar D^0)$. Using the BABAR-fitted central normalization $K_{\rm dressed}=0.382$, the corresponding dressed cross sections are $(24.6,14.6)~\mu{\rm b}$ in Pb-Pb UPCs and $(2.41,1.43)~\rm nb$ at the EIC. The charged-to-neutral ratios are directly accessible observables, whereas the dressed-to-bare ratios provide theoretical diagnostics of the final-state interaction. The near-threshold features are controlled by the common $\gamma\gamma\to D\bar D$ dynamics and are rather independent of the external
photon source.
\end{abstract}


\maketitle

\vspace{0.5cm}

\section{Introduction}

One of the central questions in hadron physics is how the exotic heavy-quark states are organized. Since the discovery of the $X(3872)$ by Belle~\cite{x3872_discovery}, many charmoniumlike states, usually denoted collectively as the $XYZ$ states, have been observed at $B$ factories, the Tevatron, and the LHC~\cite{exotics_review}. Their spectrum cannot be described by conventional charmonium alone. Compact tetraquarks~\cite{tetraquark_review}, hadronic molecules~\cite{hadronic_molecules_review}, and mixtures of these configurations are all possible, and at present there is no general experimental criterion that separates them. Production and decay observables must therefore be studied together with the mass spectrum.

A particularly simple molecular candidate is the scalar $D\bar D$ state, often called $X(3700)$. It appears in coupled-channel calculations of the $D^+D^-$ and $D^0\bar D^0$ interaction~\cite{x3700_detection,gamermann2007}, where an $SU(4)$-extended chiral contact interaction with explicit $SU(4)$ breaking generates a state close to the lowest open-charm threshold. Its predicted mass lies a few MeV below $2m_{D^0}\simeq3730$~MeV. This makes the threshold region itself, rather than an isolated narrow peak far above it, the natural place to search for the state. Evidence for related scalar dynamics has also been found in lattice-QCD calculations of the coupled $D\bar D$--$D_s\bar D_s$ system~\cite{lattice_ddbar} and in a dispersive analysis of Belle data on $\gamma\gamma\to D\bar D$~\cite{deineka_ddbarconfirmation}.

Experimentally, finding evidence for such a state is challenging because a bound or virtual state below the lightest open-charm threshold does not necessarily produce a conventional resonance peak in the measured $D\bar D$ spectrum. Its presence may instead be inferred through threshold enhancements, cusps, or correlated distortions between the charged and neutral $D\bar D$ channels~\cite{hadronic_molecules_review,deineka_ddbarconfirmation,guo_threshold_cusps}. This makes production mechanisms with clean quantum numbers and controllable kinematics especially valuable for testing the $X(3700)$ scenario.

Two-photon fusion provides a direct probe of this system. In the helicity-zero configuration, the $\gamma\gamma$ state has the quantum numbers needed to couple to a $J^{PC}=0^{++}$ $D\bar D$ pair~\cite{budnev1975}. Moreover, the electromagnetic initial state avoids leading single-photon and Pomeron contributions. The elementary process has been measured by Belle and BABAR~\cite{belle_gammagamma_ddbar,babar_gammagamma_ddbar,belle_ddbar,babar_ddbar}, so its near-threshold line shape can be confronted with data before it is used as input in the equivalent photon approximation in a high-energy collision.

At high energies the same subprocess can be studied with the intense fluxes of quasireal photons generated by charged projectiles. In ultraperipheral heavy-ion collisions at the LHC~\cite{upc_review,baltz_upc,klein_steinberg_upc}, the nuclei pass each other at impact parameters larger than their combined radii, suppressing hadronic overlap while retaining the $Z^4$ enhancement of the two-photon luminosity. The EIC offers a complementary asymmetric system~\cite{eic_review,eic_yellow_report}: a photon emitted by the electron interacts with the coherent photon field of the nucleus, and the absence of a strong-interaction radius for the electron changes both the accessible rapidity range and the treatment of impact parameters. These reactions therefore provide a way to test the same $D\bar D$ final-state interaction inferred from $B$-factory data in different photon-induced environments. Charmonium production in these environments has already been considered in Refs.~\cite{bertulani_eic,babiarz2023}.

In this work we ask how the $X(3700)$ coupled-channel scenario would appear in near-threshold $D\bar D$ production in Pb-Pb UPCs at $\sqrt{s_{NN}}=5.5$~TeV and in $e$-Au collisions with $E_e=18$~GeV and $E_A/A=100$~GeV. The calculation extends Ref.~\cite{sobrinho:2024tre} in three directions. First, the neutral $D^0\bar D^0$ production amplitude is included explicitly together with the charged channel. Second, both amplitudes are dressed consistently with the same two-channel Bethe-Salpeter equation~\cite{x3700_detection,gamermann2007}. Third, UPC and EIC observables are evaluated in a common equivalent-photon framework. The charged amplitude contains the scalar-QED terms and anomalous $D^*$ exchange, whereas the neutral amplitude is generated only by $D^{*0}$ exchange. For the photon spectra we use the electron flux proposed in Ref.~\cite{budnev1975} and a hard-sphere$\times$Yukawa nuclear form factor, with monopole and Woods-Saxon profiles used as checks.

We adopt the interaction kernel and reference parameters of Refs.~\cite{x3700_detection,gamermann2007} and study the pole structure, the charged and neutral continuum line shapes, and their convolution with UPC and EIC photon fluxes. The charged-to-neutral ratios are directly accessible observables, while dressed-to-bare ratios are used as theoretical diagnostics of the final-state interaction.

In this coupled-channel approach, the $X(3700)$ is neither introduced as an elementary field nor identified with a static state in a single fixed channel, such as $D^+D^-$ or $D^0\bar D^0$. Instead, it is generated dynamically as a pole of the coupled-channel $D\bar D$ scattering amplitude, with its charged and neutral components encoded in the corresponding pole residues~\cite{gamermann2007,x3700_detection,branz2011}. Our main results therefore concern the dressed $D\bar D$ continuum. A separate Low-formula estimate of direct bound-state production is provided only for comparison.

The paper is organized as follows. In Sec.~\ref{sec:production} we derive the charged and neutral production amplitudes and their $S$-wave cross sections. In Sec.~\ref{sec:fsi} we introduce the coupled-channel interaction and construct the dressed amplitudes. The equivalent-photon calculation for UPC and EIC kinematics is given in Sec.~\ref{sec:epa}. We present and discuss the numerical results in Sec.~\ref{sec:results}, and summarize our conclusions in Sec.~\ref{sec:conclusions}. Details of the partial-wave projection and photon spectra are collected in the appendices.
 \section{Elementary $\gamma\gamma\to D\bar{D}$ production}
\label{sec:production}

\subsection{Tree-level charged and neutral amplitudes}

We begin with the charged subprocess $\gamma\gamma\to D^+D^-$. At tree level, its scalar-QED and anomalous electromagnetic interactions follow from the Lagrangians~\cite{guo_lagrangian}

\begin{equation}
    \mathcal{L}_{D D \gamma} = (\partial_\mu D^+ + ie A_\mu D^+)(\partial^\mu D^- - ie A^\mu D^-) \, ,
\end{equation}
and
\begin{equation}
    \mathcal{L}_{DD^{*}\gamma} = -ig_{\gamma D^{+}D^{*-}} F_{\mu\nu}\epsilon^{\mu\nu\alpha\beta}(D^{*-}_\alpha\overset{\leftrightarrow}{\partial_\beta}D^{+} + D^-\overset{\leftrightarrow}{\partial_\beta}D^{*+}_{\alpha}) \, ,
\end{equation}
where $F_{\mu\nu}$ is the electromagnetic field tensor, $A_\mu$ is the photon field, $D^{(*)\pm}$ are the charged $D$ meson fields, and $g_{\gamma D^{+}D^{*-}}$ is the coupling constant for the $\gamma D^{+}D^{*-}$ vertex.

\begin{figure}[!h]
\centering
\includegraphics[width=0.95\linewidth]{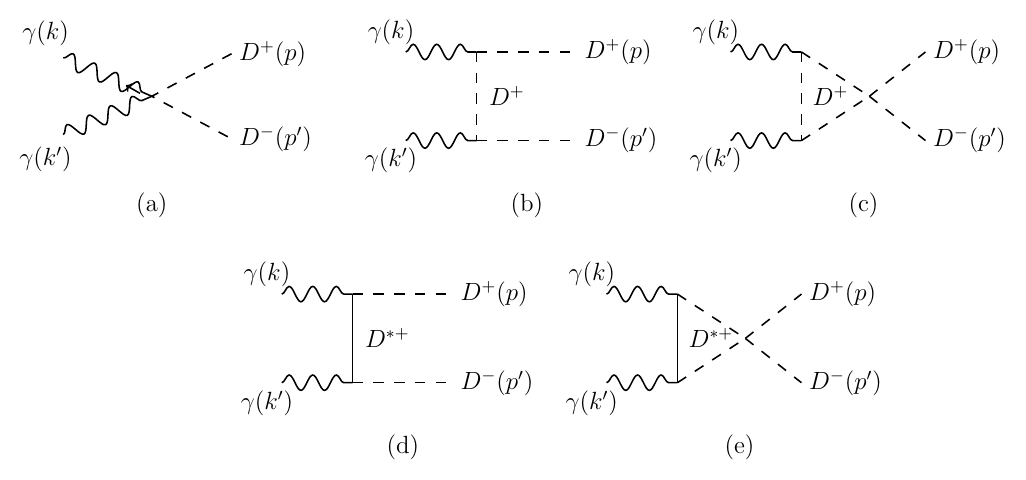}
\caption{Tree-level diagrams for $\gamma\gamma \to D^+D^-$: (a) contact term, (b,c) charged-$D$ exchange in the $t$ and $u$ channels, and (d,e) $D^*$ exchange in the $t$ and $u$ channels.}
\label{fig:feynman_charged_all}
\end{figure}

The diagrams contributing to the lowest order amplitude are shown in Fig.~\ref{fig:feynman_charged_all}. The first three are the scalar-QED contact and charged-$D$ exchange terms, while the last two describe anomalous $D^*$ exchange in the $t$ and $u$ channels.

\vskip5mm

The total amplitude is
\begin{equation}
  iM =  iM_{(a)} +  iM_{(b)} +  iM_{(c)}  + iM_{(d)} +  iM_{(e)} \, ,
\label{tamp}
\end{equation}
where
\begin{align}
iM_{(a)} &= 2ie^2g_{\mu\nu}
            \varepsilon^{\mu}(k) \varepsilon^{\nu}(k') \, ,\\
iM_{(b)} &= \varepsilon^{\mu}(k)ie (-2p_\mu + k_\mu)
\frac{i}{(k-p)^2-m_D^2}ie (2p'_\nu - k'_\nu)
\varepsilon^{\nu}(k') \, ,\\
iM_{(c)} &= \varepsilon^{\mu}(k')ie(-2p_\mu + k'_\mu)
\frac{i}{(k'-p)^2-m_D^2}ie(2p'_\nu - k_\nu)
\varepsilon^{\nu}(k) \, , \\
iM_{(d)} &= \varepsilon_{\mu}(k)[2g_{\gamma,+}\epsilon^{\sigma\mu\alpha\rho}k_{\sigma}(k_\rho - 2p_\rho)]\left[\frac{-i(g_{\alpha\beta} - \frac{(k-p)_\alpha (k-p)_\beta}{m_{D^*}^2})}{(k-p)^2-m_{D^*}^2}\right][2g_{\gamma,+}\epsilon^{\delta\nu\beta\lambda}k'_{\delta}(-k'_\lambda + 2p'_\lambda)]\varepsilon_{\nu}(k')\, , \\
iM_{(e)} &= \varepsilon_{\mu}(k')[2g_{\gamma,+}\epsilon^{\sigma\mu\alpha\rho}k'_{\sigma}(k'_\rho - 2p_\rho)]\left[\frac{-i(g_{\alpha\beta} - \frac{(k'-p)_\alpha (k'-p)_\beta}{m_{D^*}^2})}{(k'-p)^2-m_{D^*}^2}\right][2g_{\gamma,+}\epsilon^{\delta\nu\beta\lambda}k_{\delta}(-k_\lambda + 2p'_\lambda)]\varepsilon_{\nu}(k)\, .
 \end{align}
Here $g_{\gamma,+}\equiv g_{\gamma D^{+}D^{*-}}=-0.035$~GeV$^{-1}$ denotes the charged electromagnetic $\gamma DD^*$ coupling~\cite{guo_lagrangian}.

\begin{figure}[!h]
\centering
\includegraphics[width=0.72\linewidth]{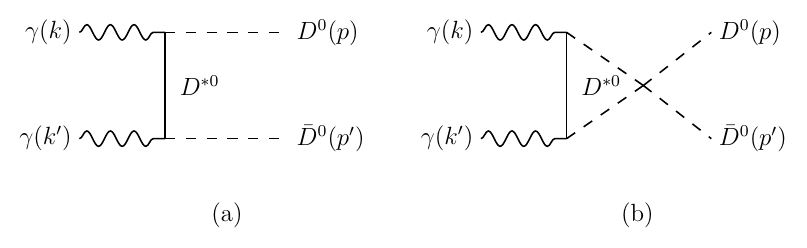}
\caption{Tree-level diagrams for $\gamma\gamma \to D^0\bar{D}^0$ with $D^{*0}$ exchange in the $t$ and $u$ channels.}
\label{fig:feynman_dstar0}
\end{figure}

We now extend the calculation of Ref.~\cite{sobrinho:2024tre} and include the neutral channel. Because the $D^0$ carries no electric charge, there is no scalar-QED contact term or $D^0$ exchange. The process $\gamma\gamma\to D^0\bar D^0$ is determined through the anomalous interaction~\cite{guo_lagrangian}
\begin{equation}
    \mathcal{L}_{D^0D^{*0}\gamma} = -ig_{\gamma D^{0}D^{*0}} F_{\mu\nu}\epsilon^{\mu\nu\alpha\beta}(\bar{D}^{*0}_\alpha\overset{\leftrightarrow}{\partial_\beta}D^{0} + \bar{D}^0\overset{\leftrightarrow}{\partial_\beta}D^{*0}_{\alpha}) \, ,
\end{equation}
which produces the diagrams shown in Fig.~\ref{fig:feynman_dstar0}.

The total amplitude for the $\gamma\gamma \to D^0\bar{D}^0$ process is
\begin{equation}
  iM_{neutral} = iM_{(a)}^{(0)} + iM_{(b)}^{(0)} \, ,
\label{tamp_neutral}
\end{equation}
with
\begin{align}
iM_{(a)}^{(0)} &= \varepsilon_{\mu}(k)[2g_{\gamma,0}\epsilon^{\sigma\mu\alpha\rho}k_{\sigma}(k_\rho - 2p_\rho)]\left[\frac{-i(g_{\alpha\beta} - \frac{(k-p)_\alpha (k-p)_\beta}{m_{D^{*0}}^2})}{(k-p)^2-m_{D^{*0}}^2}\right][2g_{\gamma,0}\epsilon^{\delta\nu\beta\lambda}k'_{\delta}(-k'_\lambda + 2p'_\lambda)]\varepsilon_{\nu}(k')\, , \\
iM_{(b)}^{(0)} &= \varepsilon_{\mu}(k')[2g_{\gamma,0}\epsilon^{\sigma\mu\alpha\rho}k'_{\sigma}(k'_\rho - 2p_\rho)]\left[\frac{-i(g_{\alpha\beta} - \frac{(k'-p)_\alpha (k'-p)_\beta}{m_{D^{*0}}^2})}{(k'-p)^2-m_{D^{*0}}^2}\right][2g_{\gamma,0}\epsilon^{\delta\nu\beta\lambda}k_{\delta}(-k_\lambda + 2p'_\lambda)]\varepsilon_{\nu}(k)\, ,
\end{align}
Here $g_{\gamma,0}\equiv g_{\gamma D^{0}D^{*0}}=0.142$~GeV$^{-1}$ denotes the neutral electromagnetic $\gamma DD^*$ coupling~\cite{guo_lagrangian}.

To consider final-state interactions, we define the charge-basis vector
\begin{equation}
\mathbf{M}_{\rm ch}
=
\begin{pmatrix}
\mathcal M_{+-}\\
\mathcal M_{00}
\end{pmatrix}
\equiv
\begin{pmatrix}
\mathcal M(\gamma\gamma\to D^+D^-)\\
\mathcal M(\gamma\gamma\to D^0\bar D^0)
\end{pmatrix} .
\label{eq:charge_basis_vector}
\end{equation}
For the single-meson isodoublet we use the phase convention
\begin{equation}
|D\rangle_{1/2}=
\begin{pmatrix}
|D^+\rangle\\
-|D^0\rangle
\end{pmatrix},
\qquad
|\bar D\rangle_{1/2}=
\begin{pmatrix}
|\bar D^0\rangle\\
|D^-\rangle
\end{pmatrix},
\label{eq:isodoublet_convention}
\end{equation}
so that for $I_3=0$ two-meson states one gets
\begin{equation}
|I=0\rangle=\frac{|D^+D^-\rangle+|D^0\bar D^0\rangle}{\sqrt2},
\qquad
|I=1\rangle=\frac{|D^+D^-\rangle-|D^0\bar D^0\rangle}{\sqrt2} .
\label{eq:isospin_states_ddbar}
\end{equation}
Equivalently, amplitudes are related by
\begin{equation}
\begin{pmatrix}
\mathcal M_{I=0}\\
\mathcal M_{I=1}
\end{pmatrix}
=
\frac{1}{\sqrt2}
\begin{pmatrix}
1 & 1\\
1 & -1
\end{pmatrix}
\begin{pmatrix}
\mathcal M_{+-}\\
\mathcal M_{00}
\end{pmatrix}.
\label{eq:basis_transform_amplitudes}
\end{equation}
Thus the isospin amplitudes are coherent sums: they are added before the modulus square is taken. In the threshold region, where the $I=0$ component dominates, the shorthand notation
\begin{equation}
  iM(\gamma\gamma\to D\bar{D}) = \frac{i}{\sqrt2}\left[M(\gamma\gamma\to D^+D^-) + M(\gamma\gamma\to D^0\bar{D}^0)\right] \, .
  \label{tamp_total}
\end{equation}
represents the $I=0$ projection. In the approach followed in Sec.~\ref{sec:fsi}, however, we retain the physical charge basis $(D^+D^-,D^0\bar D^0)$ throughout the dressing and use the isospin basis only for the interpretation of the results.

It is useful to establish the gauge properties before discussing finite-size effects. The point-like tree-level amplitudes, where by "point-like" we mean considering mesons without internal structure, are gauge invariant. In particular, the charged amplitude satisfies
\begin{equation}
k_\mu M^{\mu\nu}_{\rm ch}=0\,,\qquad k'_\nu M^{\mu\nu}_{\rm ch}=0\,.
\end{equation}
The Ward identities follow from the cancellation between the scalar-QED contact and $D$-exchange terms, while each anomalous $D^*$ contribution is transverse by itself. The neutral amplitude is also transverse because it contains only anomalous $D^{*0}$ exchange. Gauge invariance is therefore already present in the point-like theory.

\subsection{Finite-size effects and normalization}

Finite-size effects may be represented by a monopole factor at each hadronic electromagnetic vertex~\cite{Kim:2015vwp,Nam:2008jy},
\begin{equation}
F_X(q^2)=\frac{\Lambda_X^2-m_X^2}{\Lambda_X^2-q^2}\,,
\qquad F_X(m_X^2)=1\, ,
\label{eq:vertex_ff_monopole}
\end{equation}
where $X\in\{D^\pm,D^{*\pm},D^{*0}\}$, $m_X$ represents its mass, $\Lambda_X$ is the cutoff parameter, and $q^2$ is the momentum of the exchanged hadron at that vertex. With this choice, and defining the Mandelstam variables $\hat{s}=(k+k')^2$, $\hat{t}=(k-p)^2$ and $\hat{u}=(k'-p)^2$, each of the diagrams is dressed as
\begin{align}
M_{(a)} &\to F_{D^\pm}(\hat t)\,F_{D^\pm}(\hat u)\,M_{(a)}\,,\nonumber\\
M_{(b)} &\to F_{D^\pm}^2(\hat t)\,M_{(b)}\,,\qquad
M_{(c)} \to F_{D^\pm}^2(\hat u)\,M_{(c)}\,,\nonumber\\
M_{(d)} &\to F_{D^{*\pm}}^2(\hat t)\,M_{(d)}\,,\qquad
M_{(e)} \to F_{D^{*\pm}}^2(\hat u)\,M_{(e)}\,,\nonumber\\
M_{(a)}^{(0)} &\to F_{D^{*0}}^2(\hat t)\,M_{(a)}^{(0)}\,,\qquad
M_{(b)}^{(0)} \to F_{D^{*0}}^2(\hat u)\,M_{(b)}^{(0)}\,.
\label{eq:ff_all_diagrams}
\end{align}
Here $m_{D^\pm}$ is used for the charged-$D$ exchange pieces, $m_{D^{*\pm}}$ for charged-$D^*$ exchange, and $m_{D^{*0}}$ for neutral-$D^*$ exchange, as required by the exchanged hadron in each diagram. We consider the product of two form factors for the contact term as a prescription to suppress the high-energy growth of the amplitude, since the contact term does not have a momentum dependence in the point-like description~\cite{Kim:2015vwp,Nam:2008jy}. 

Using the convention in which $\varepsilon_\mu$ annihilates an incoming photon and $\varepsilon^*_\mu$ creates one, the amplitudes can be written as $M_{(x)} = M_{(x)}^{\mu\nu}\varepsilon_\mu(k)\varepsilon_\nu(k')$. At the point-like level, the charged scalar-QED subset $(a)+(b)+(c)$ is gauge invariant by itself,
\begin{equation}
k_\mu\,\bigl(M_{(a)}^{\mu\nu}+M_{(b)}^{\mu\nu}+M_{(c)}^{\mu\nu}\bigr)=0\, ,
\qquad
k'_\nu\,\bigl(M_{(a)}^{\mu\nu}+M_{(b)}^{\mu\nu}+M_{(c)}^{\mu\nu}\bigr)=0\,.
\label{eq:ward_pointlike_scalarqed}
\end{equation}
After vertex dressing, however, each term is multiplied by a different kinematic function ($\hat t$ or $\hat u$), and the cancellation no longer holds exactly, as is well known in phenomenological descriptions of electromagnetic hadron production~\cite{Ohta:1989ji,Haberzettl:1997jg,Davidson:2001rk,Haberzettl:2006bn}. Defining
\begin{equation}
f_a\equiv F_{D^\pm}(\hat t)F_{D^\pm}(\hat u),\qquad
f_t\equiv F_{D^\pm}^2(\hat t),\qquad
f_u\equiv F_{D^\pm}^2(\hat u),
\end{equation}
one finds, using Eq.~\eqref{eq:ward_pointlike_scalarqed},
\begin{equation}
k_\mu\,\widetilde M_{\rm ch}^{\mu\nu}
= k_\mu\left(f_a\,M_{(a)}^{\mu\nu}+f_t\,M_{(b)}^{\mu\nu}+f_u\,M_{(c)}^{\mu\nu}\right)
=
-(f_a-f_t)\,k_\mu M_{(b)}^{\mu\nu}
-
(f_a-f_u)\,k_\mu M_{(c)}^{\mu\nu}\neq 0
\quad (\hat t\neq \hat u)\,.
\label{eq:ward_broken_ff}
\end{equation}
Thus a diagram-by-diagram insertion of form factors spoils the exact cancellation in the charged scalar-QED sector. The anomalous $D^*$ terms remain individually transverse because the $\gamma DD^*$ vertex is constructed from $F_{\mu\nu}$.

Near threshold this problem simplifies because $\hat t$ and $\hat u$ are close and vary only weakly,
\begin{equation}
\hat t=\bar q^2+\Delta\,,\qquad
\hat u=\bar q^2-\Delta\,,\qquad
\Delta\propto \beta_D\cos\theta\,.
\end{equation}
where $\beta_D = \sqrt{1-4m_D^2/\hat{s}}$ is the center-of-mass velocity of the outgoing $D$ mesons, $\bar q^2=(\hat t+\hat u)/2$ and $\Delta=(\hat t-\hat u)/2$.
Expanding Eq.~\eqref{eq:vertex_ff_monopole} around $\Delta$ gives
\begin{equation}
F_X(\hat t)=F_X(\bar q^2)+\mathcal O(\Delta),
\qquad
F_X(\hat u)=F_X(\bar q^2)+\mathcal O(\Delta),
\end{equation}
so that, at leading order in $\beta_D$,
\begin{equation}
F_X(\hat t)F_X(\hat u)
\simeq
F_X^2(\hat t)
\simeq
F_X^2(\hat u)
\simeq
F_X^2(\bar q^2)\,.
\label{eq:ff_factorization_threshold}
\end{equation}
Hence the low-energy amplitude factorizes as
\begin{align}
M_{\rm ch}^{\rm FF}
&\simeq
F_{D^\pm}^2(\bar q^2)\,[M_{(a)}+M_{(b)}+M_{(c)}]
+
F_{D^{*\pm}}^2(\bar q^2)\,[M_{(d)}+M_{(e)}],\nonumber\\
M_{\rm n}^{\rm FF}
&\simeq
F_{D^{*0}}^2(\bar q^2)\,[M_{(a)}^{(0)}+M_{(b)}^{(0)}].
\label{eq:ff_factorized_amplitudes}
\end{align}
The anomalous $D^*$-exchange contributions are individually transverse. Multiplication by scalar functions of $\hat t$ or $\hat u$ does not alter this property, so the anomalous charged contributions and the complete neutral amplitude remain exactly transverse after the form factors are introduced. Threshold factorization is needed only to recover the charged scalar-QED cancellation approximately:
\begin{equation}
k_\mu M_{\rm ch}^{\mu\nu,\rm FF}=\mathcal O(\beta_D),
\qquad
k'_\nu M_{\rm ch}^{\mu\nu,\rm FF}=\mathcal O(\beta_D),.
\end{equation}
The neutral-channel Ward identities, by contrast, hold exactly and do not rely on the near-threshold approximation.

This threshold factorization determines how finite-size effects are treated numerically. We do not multiply the individual diagrams by different form factors. Instead, we use Eq.~\eqref{eq:ff_factorized_amplitudes} to motivate an approximately multiplicative uncertainty for energies close to threshold: the omitted short-distance structure changes mainly the overall scale, while the gauge-invariant kinematic dependence and the channel-dependent FSI pattern are retained at leading order.

\subsection{$S$-wave subprocess cross section}

At the energies close to threshold considered here, the partial wave with zero orbital angular momentum (the $S$ wave) dominates the interaction between the $D$ and $\bar D$ mesons. Indeed, as shown in Refs.~\cite{hadronic_molecules_review,x3700_detection,gamermann2007}, the $X(3700)$ is generated from the $S$-wave interaction of the $D\bar D$ system. With this in mind, we project the charged and neutral amplitudes in Eqs.~\eqref{tamp} and~\eqref{tamp_neutral} onto the $S$ wave by integrating over the center-of-mass scattering angle:
\begin{equation}
  \mathcal{M}_0(\hat{s}) = \frac{1}{2}\int_{-1}^{1} d\cos\theta\, \mathcal{M}(\hat{s},\cos\theta) \, ,
  \label{eq:swave_proj}
\end{equation}
where the first subscript $0$ indicates the projection onto the $S$ wave. For the channel-specific amplitudes below, the second subscript follows the charge-basis notation of Eq.~\eqref{eq:charge_basis_vector}: $+-$ for $D^+D^-$ and $00$ for $D^0\bar D^0$. This integral can be performed analytically for each diagram (for details, see Appendix~\ref{app:projection}).
For the charged channel the result obtained is
\begin{equation}
  \mathcal{M}_{0,+-}(\hat{s}) = 4i\left[\frac{2e^2m_{D^\pm}^2\,\mathrm{arctanh}(\beta_{D^\pm}) - 4g_{\gamma,+}^2m_{D^{*\pm}}^2\hat{s}\,\mathrm{arccoth}(z^*)}{\sqrt{\hat{s}(\hat{s}-4m_{D^\pm}^2)}} + 2g_{\gamma,+}^2\hat{s}\right] \, ,
  \label{eq:M0_charged}
\end{equation}
where $\beta_{D^\pm} = \sqrt{1-4m_{D^\pm}^2/\hat{s}}$ is the center-of-mass velocity of the outgoing $D$ mesons and $z^* = (2m_{D^{*\pm}}^2 - 2m_{D^\pm}^2 + \hat{s})/\sqrt{\hat{s}(\hat{s}-4m_{D^\pm}^2)}$. For the neutral channel, where only $D^{*0}$ exchange contributes, we obtain
\begin{equation}
  \mathcal{M}_{0,00}(\hat{s}) = -4ig_{\gamma,0}^2\hat{s}\left[\frac{4m_{D^{*0}}^2\,\mathrm{arccoth}(z_0)}{\sqrt{\hat{s}(\hat{s}-4m_{D^0}^2)}} - 2\right] \, ,
  \label{eq:M0_neutral}
\end{equation}
with $z_0 = (2m_{D^{*0}}^2 - 2m_{D^0}^2 + \hat{s})/\sqrt{\hat{s}(\hat{s}-4m_{D^0}^2)}$.

The charged and neutral final states are observed separately. Their subprocess cross sections are therefore calculated channel by channel,
\begin{equation}
  \hat{\sigma}_i(\hat{s}) = \frac{\beta_i}{32\pi\hat{s}} |\mathcal{M}_{0,i}(\hat{s})|^2 \, ,
  \qquad i\in\{+-,00\},
  \label{eq:sigmahat}
\end{equation}
where $\beta_{+-}=\sqrt{1-4m_{D^\pm}^2/\hat{s}}$ and $\beta_{00}=\sqrt{1-4m_{D^0}^2/\hat{s}}$. The coherent combination $(\mathcal{M}_{0,+-}+\mathcal{M}_{0,00})/\sqrt{2}$ is the $I=0$ amplitude defined in Eq.~\eqref{tamp_total}; it is not used to combine the two distinct experimental final states.

The meson masses used are $m_{D^\pm} = 1.86965$ GeV, $m_{D^0} = 1.86483$ GeV, $m_{D^{*\pm}} = 2.01026$ GeV, and $m_{D^{*0}} = 2.00685$ GeV \cite{PDG2024}. With these values, the $D^+D^-$ threshold is at $W_{\rm thr}^{\rm ch} = 2m_{D^\pm} = 3.7393$ GeV and the $D^0\bar{D}^0$ threshold is at $W_{\rm thr}^{0} = 2m_{D^0} = 3.7297$ GeV. The electromagnetic coupling is $e = \sqrt{4\pi\alpha_{\rm em}}$ with $\alpha_{\rm em} \simeq 1/137$.
 \section{Final-state interaction}
\label{sec:fsi}

The amplitudes derived in Sec.~\ref{sec:production} produce the meson pair, but near threshold the outgoing $D$ and $\bar D$ can rescatter before they are observed. We describe this final-state interaction through the $T$-matrix obtained from the coupled-channel Bethe-Salpeter equation of Ref.~\cite{gamermann2007}. For completeness, we summarize the main ingredients of that approach below.

\subsection{Coupled-channel dynamics}

We work with the charged pair ($D^+D^-$, channel~1) and the neutral pair ($D^0\bar D^0$, channel~2). In the on-shell factorized approximation of the Bethe-Salpeter equation~\cite{oset_chiral_unitary,oset_ramos,oller_meissner,hyodo_jido}, the integral equation
\begin{equation}
	T = V + V\,G\,T \, ,
	\label{eq:bs_equation}
\end{equation}
reduces to an algebraic equation involving, in this case, $2\times 2$ matrices: 
\begin{equation}
	T(\hat s) = \bigl[I - V(\hat s)\,G(\hat s)\bigr]^{-1}\!V(\hat s) \, ,
	\label{eq:t_matrix}
\end{equation}
where $I$ is the identity matrix, $V(\hat s)$ is the interaction kernel, and $G(\hat s)=\mathrm{diag}(G_1,G_2)$ contains the two-meson loops. Throughout this subprocess calculation, $W=\sqrt{\hat{s}}$ denotes the $D\bar D$ invariant mass.

Following Ref.~\cite{gamermann2007}, the kernel in the physical charge basis is obtained from the $SU(4)$-broken extension of the lowest-order chiral contact interaction. Its matrix elements are
\begin{align}
	V_{11}(\hat{s}) &= \frac{-1}{3f_D^2}\left[\psi_3\!\left(\frac{3}{2}\hat{s} - 2m_{D^\pm}^2\right) + 2m_{D^\pm}^2\right] \, ,
	\label{eq:V11}\\[4pt]
	V_{22}(\hat{s}) &= \frac{-1}{3f_D^2}\left[\psi_3\!\left(\frac{3}{2}\hat{s} - 2m_{D^0}^2\right) + 2m_{D^0}^2\right] \, ,
	\label{eq:V22}\\[4pt]
	V_{12}(\hat{s}) &= V_{21}(\hat{s}) = \frac{-\hat{s}}{4f_D^2} \, ,
	\label{eq:V12}
\end{align}
with $f_D=165$~MeV and $\psi_3 = \frac{1}{3} + \frac{2}{3}\left(\frac{m_\sigma}{m_{J/\psi}}\right)^{\!2} \approx 0.378$, where $m_\sigma=800$~MeV and $m_{J/\psi}=3097$~MeV~\cite{PDG2024}, accounts for the $SU(4)$ breaking in the model. The off-diagonal element $V_{12}$ converts charged pairs into neutral pairs and vice versa, and therefore plays an essential role in the final-state interaction.

To solve Eq.~\eqref{eq:t_matrix}, we also need the loop function for two pseudoscalar mesons of masses $m_1$ and $m_2$. Within the dimensional regularization scheme we have~\cite{gamermann2007}
\begin{align}
	G(\hat{s}) = \frac{1}{16\pi^2}\Bigg[&\alpha_H + \ln\frac{m_1^2}{\mu^2}
	+ \frac{m_2^2 - m_1^2 + \hat{s}}{2\hat{s}}\ln\frac{m_2^2}{m_1^2}
	\nonumber \\
	&+ \frac{p(\hat{s})}{\sqrt{\hat{s}}}\left(
	   \ln\frac{\hat{s} - m_2^2 + m_1^2 + 2p(\hat{s})\sqrt{\hat{s}}}{-\hat{s} + m_2^2 - m_1^2 + 2p(\hat{s})\sqrt{\hat{s}}}
	   + \ln\frac{\hat{s} + m_2^2 - m_1^2 + 2p(\hat{s})\sqrt{\hat{s}}}{-\hat{s} - m_2^2 + m_1^2 + 2p(\hat{s})\sqrt{\hat{s}}}
	\right)
	\Bigg] \, ,
	\label{eq:loop_general}
\end{align}
where the modulus of the center-of-mass momentum of the particles is
\begin{equation}
	p(\hat{s}) = \frac{\sqrt{\bigl[\hat{s}-(m_1+m_2)^2\bigr]\bigl[\hat{s}-(m_1-m_2)^2\bigr]}}{2\sqrt{\hat{s}}} \, .
	\label{eq:pcm}
\end{equation}
In Eq.~\eqref{eq:loop_general}, the subtraction constant $\alpha_H$ and scale $\mu$ parametrize the ultraviolet behavior of the loop. Because a change in $\mu$ is compensated by a change in $\alpha_H$, they represent one effective parameter. We use the values of Ref.~\cite{gamermann2007}: $\alpha_H=-1.3$ and $\mu=1500$~MeV.

For $\hat{s}$ below threshold $(m_1+m_2)^2$, the momentum is purely imaginary, $p = i|p|$ and the logarithmic terms in Eq.~\eqref{eq:loop_general} combine into a real result. For $\hat{s}$ above threshold, $p$ is real and $G(\hat{s})$ develops the standard unitarity imaginary part \cite{oset_chiral_unitary},
\begin{equation}
	\mathrm{Im}\,G(\hat{s}) = -\frac{p(\hat{s})}{8\pi\sqrt{\hat{s}}} \, ,\qquad \hat{s} > (m_1+m_2)^2 \, ,
	\label{eq:ImG}
\end{equation}
which is the cut required by the optical theorem.

The analytic continuation of Eq.~\eqref{eq:t_matrix} to the complex energy plane, using the interaction of Refs.~\cite{gamermann2007,x3700_detection}, produces a pole at
\begin{equation}
 W_X=M_X-\frac{i}{2}\Gamma_X
 =3.7162~{\rm GeV},\qquad \Gamma_X=0 ,
 \label{eq:x3700_pole_mass_width}
\end{equation}
which is denoted as $X(3700)$. Since the pole position lies below both the $D^0\bar D^0$ and $D^+D^-$ thresholds, it is a bound state. This result is compatible with the near-threshold enhancement found in the dispersive analysis of Belle data~\cite{deineka_ddbarconfirmation}.

The couplings of the pole found to the two physical channels are obtained from the residues of the $T$-matrix: for energies $W$ close to the pole position $W_X$, with $\hat{s}_X=W_X^2$, we have
\begin{equation}
 T_{ij}(\hat{s})\simeq \frac{g_i g_j}{\hat{s}-\hat{s}_X},
 \qquad
 g_i g_j=\lim_{\hat{s}\to \hat{s}_X}(\hat{s}-\hat{s}_X)T_{ij}(\hat{s}) .
 \label{eq:pole_residue_couplings}
\end{equation}
To distinguish the pole couplings from the electromagnetic couplings $g_{\gamma,+}$ and $g_{\gamma,0}$, we denote the couplings of the $X(3700)$ pole to the charged and neutral channels by $g_{X,+-}$ and $g_{X,00}$, respectively:
\begin{equation}
 g_{X,+-}=8.512~{\rm GeV},\qquad
 g_{X,00}=8.604~{\rm GeV}.
 \label{eq:x3700_channel_couplings}
\end{equation}

\subsection{$X(3700)$ threshold structure and dressed production amplitudes}

\begin{figure}[htbp]                                                              
\centering                                                                        
\includegraphics[width=0.95\linewidth]{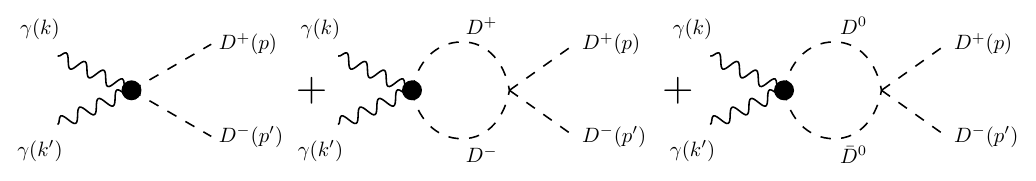} \\ 
\vskip0.5cm
\includegraphics[width=0.95\linewidth]{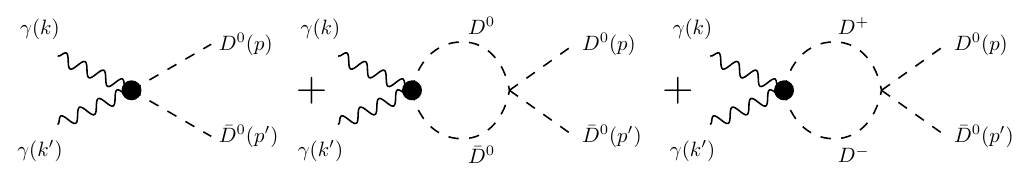}     
\caption{Diagrams contributing to $\gamma\gamma\to D^+D^-$ (top)   
and $D^0\bar D^0$ (bottom), including final-state interactions.}
\label{fig:fsi_dressing}                                                          
\end{figure}

We next implement the final-state interaction in $\gamma\gamma \to D\bar D$. The dressed amplitude for final channel $i$ is~\cite{oset_chiral_unitary}
\begin{equation}
	\mathcal{M}_i^{\rm dressed}(\hat{s}) = \mathcal{M}_i^{\rm bare}(\hat{s})
	+ \sum_{j=1}^{2} \mathcal{M}_j^{\rm bare}(\hat{s})\,G_j(\hat{s})\,T_{ji}(\hat{s}) \, ,
	\label{eq:dressed_amplitude}
\end{equation}
where $\mathcal M_j^{\rm bare}$ is the tree-level source, $G_j$ propagates the intermediate pair, and $T_{ji}$ converts the former into the observed channel. Thus
\begin{align}
	\mathcal{M}_{\rm ch}^{\rm dressed} &= \mathcal{M}_{\rm ch}^{\rm bare}
	+ \mathcal{M}_{\rm ch}^{\rm bare}\,G_{\rm ch}\,T_{11}
	+ \mathcal{M}_{\rm n}^{\rm bare}\,G_{\rm n}\,T_{12} \, ,
	\label{eq:dressed_charged}\\[4pt]
	\mathcal{M}_{\rm n}^{\rm dressed} &= \mathcal{M}_{\rm n}^{\rm bare}
	+ \mathcal{M}_{\rm ch}^{\rm bare}\,G_{\rm ch}\,T_{21}
	+ \mathcal{M}_{\rm n}^{\rm bare}\,G_{\rm n}\,T_{22} \, .
	\label{eq:dressed_neutral}
\end{align}
Figure~\ref{fig:fsi_dressing} represents the content of these equations diagrammatically. As can be seen, one of the contributions involves elastic transitions while the other depends on the inelastic transition between the charged and neutral channels.

%
%

The channel-changing terms in Fig.~\ref{fig:fsi_dressing} are important because $|g_{\gamma,0}|^2\gg |g_{\gamma,+}|^2$. The neutral bare production amplitude therefore gives a sizable contribution to the charged final state through $\mathcal M_{\rm n}^{\rm bare}G_{\rm n}T_{12}$, increasing the charged cross section near threshold. Conversely, $\mathcal M_{\rm ch}^{\rm bare}G_{\rm ch}T_{21}$ partially cancels the neutral bare amplitude and suppresses the neutral cross section. These effects are shown numerically in Sec.~\ref{sec:results}.
 \section{Equivalent-photon approximation}
\label{sec:epa}

We now include the elementary process in the high-energy collision calculation. A relativistic charged particle emits quasireal photons, and the cross section for producing a final state $\mathcal B$ in $eA$ or $AA$ collisions can be written as~\cite{budnev1975,upc_review,baltz_upc,klein_steinberg_upc}
\begin{equation}
    \sigma(A B \rightarrow A \otimes \mathcal{B} \otimes B; \sqrt{s_{AB}}) =
    \int \mbox{d}\omega_A \, \mbox{d}\omega_B\,
    f_{\gamma/A}(\omega_A)\,f_{\gamma/B}(\omega_B)\,
    \hat{\sigma}(\gamma\gamma\rightarrow\mathcal{B}; W_{\gamma\gamma})
    \, ,
\label{eq:cross_section}
\end{equation}
where $f_{\gamma/A}$ and $f_{\gamma/B}$ are the two photon spectra and $W_{\gamma\gamma}=\sqrt{4\omega_A\omega_B}$. For our differential observables it is convenient to replace the two photon energies by the pair invariant mass $W$ and rapidity $Y$,
\begin{equation}
  \omega_1 = \frac{W}{2}e^{+Y} \, , \qquad \omega_2 = \frac{W}{2}e^{-Y} \, ,
  \label{eq:omega_wy}
\end{equation}
so that the Jacobian gives
\begin{equation}
  \frac{d^2\sigma}{dW\,dY} = \frac{W}{2} f_1(\omega_1)f_2(\omega_2)\,\hat{\sigma}(\gamma\gamma\to D\bar{D};W) \, .
  \label{eq:d2sigma_wy}
\end{equation}
For $e$-Au collisions, we associate $\omega_1$ with the photon emitted by the electron and $\omega_2$ with the photon emitted by the nucleus. Positive $Y$ therefore corresponds to $\omega_e>\omega_A$ in our convention.
The one-dimensional spectra follow by integrating over the other variable,
\begin{equation}
  \frac{d\sigma}{dW} = \int dY\,\frac{d^2\sigma}{dW\,dY} \, , \qquad
  \frac{d\sigma}{dY} = \int dW\,\frac{d^2\sigma}{dW\,dY} \, .
  \label{eq:projected_distributions}
\end{equation}
For the electron we use the spectrum of Ref.~\cite{budnev1975},
\begin{equation}
  f_{\gamma/e}(\omega_e) =
  \frac{\alpha_{em}}{\pi \, \omega_e} \int\frac{\mbox{d}Q^2}{Q^2}
  \left[
    \left(
      1-\frac{\omega_e}{E_e}+\frac{\omega_e^2}{2E_e^2}
    \right)-
    \left(
      1-\frac{\omega_e}{E_e}
    \right)\frac{Q^2_{min}}{Q^2}
  \right]
  \, ,
\label{eq:flux_electron}
\end{equation}
where $Q^2_{\min}=m_e^2\omega_e^2/[E_e(E_e-\omega_e)]$ and $Q^2_{\max}$ enforces the quasireal-photon condition.
For a nucleus, coherence and finite size enter through the elastic charge form factor $F_A(q)$. The corresponding spectrum is~\cite{krauss1997}
\begin{equation}
    f_{\gamma/A}(\omega_A) =
    \frac{Z^2\alpha}{\pi^2} \int \mbox{d}^2b \, \frac{1}{b^2v^2\omega_A}
    \left[
    \int \mbox{d} u \, u^2 F_A
        \left(
            \sqrt{\frac{\left(\frac{b\omega_A}{\gamma_L}\right)^2 + u^2}{b^2}}
        \right)
        \frac{1}{\left(\frac{b\omega_A}{\gamma_L}\right)^2 + u^2}
        \, J_1(u)
    \right]^2
    \, ,
\label{eq:flux_nucleus}
\end{equation}
where $u=k_\perp b$, $\gamma_L$ is the Lorentz factor of the source, and $J_1$ is a Bessel function. Our baseline is the hard-sphere$\times$Yukawa form factor~\cite{klein_hardsphere_yukawa}
\begin{equation}
  F_A^{\rm real}(q)=\frac{4\pi\rho_0}{Aq^3}\left[\sin(qR_A)-qR_A\cos(qR_A)\right]\,\frac{1}{1+a_A^2q^2} \, ,
  \label{eq:realistic_ff}
\end{equation}
with $\rho_0=3A/(4\pi R_A^3)$, so that $F_A^{\rm real}(0)=1$. The radius $R_A$ and Yukawa range $a_A$ used for Au and Pb are listed in Appendix~\ref{app:epa}. In $eA$ collisions the coherent nuclear flux is integrated over all impact parameters with the momentum-space expression in Eq.~\eqref{eq:eic_all_b_flux}; no nuclear-overlap veto is needed for the electron. In Pb-Pb collisions we impose $b>2R_A$ to remove hadronic overlap. As a check we also use the monopole form
\begin{equation}
  F_A(q^2) = \frac{\Lambda_A^2}{\Lambda_A^2 + q^2} \, ,
  \label{eq:monopole_ff}
\end{equation}
with $\Lambda_{\rm Au}=0.091$~GeV and $\Lambda_{\rm Pb}=0.088$~GeV~\cite{monopole_parameters_1,monopole_parameters_2}.
We also compare point-like, monopole, and Woods-Saxon charge profiles~\cite{klein_hardsphere_yukawa}. We use the hard-sphere$\times$Yukawa form factor as the default. The numerical comparison among the different form factors is presented in Appendix~\ref{app:epa}.
 \section{Results}
\label{sec:results}

We now present the numerical results. For UPCs we consider Pb-Pb collisions at $\sqrt{s_{NN}}=5.5$~TeV, with $\gamma_L=2750$. For the EIC we use $E_e=18$~GeV and $E_A/A=100$~GeV in $e$-Au collisions, corresponding to $\gamma_L\simeq106.6$ for the ion beam. We calculate $d\sigma/dW$, $d\sigma/dY$, the midrapidity spectrum $d^2\sigma/(dW\,dY)|_{Y=0}$, and the cross section integrated over $3.70\leq W\leq3.85$~GeV and $|Y|\leq4$. The upper mass limit excludes the additional structure near $3.86$~GeV, which is not contained in the present $S$-wave model~\cite{belle_gammagamma_ddbar,babar_gammagamma_ddbar,deineka_ddbarconfirmation}. All absolute bare and dressed cross sections quoted in the Tables are rescaled by the BABAR dressed central value $K=0.382$.

The UPC and EIC spectra are obtained from the EPA expression in Eq.~\eqref{eq:d2sigma_wy}. In that equation the subprocess cross section $\hat\sigma(\gamma\gamma\to D\bar D;W)$ is the elementary two-photon cross section defined in Eq.~\eqref{eq:sigmahat}, evaluated with either the bare amplitude or the Bethe-Salpeter dressed amplitude of Eq.~\eqref{eq:dressed_amplitude}. 

\subsection{Elementary $X(3700)$ line-shape signature}

\begin{figure}[htbp]
\centering
\includegraphics[width=1\linewidth]{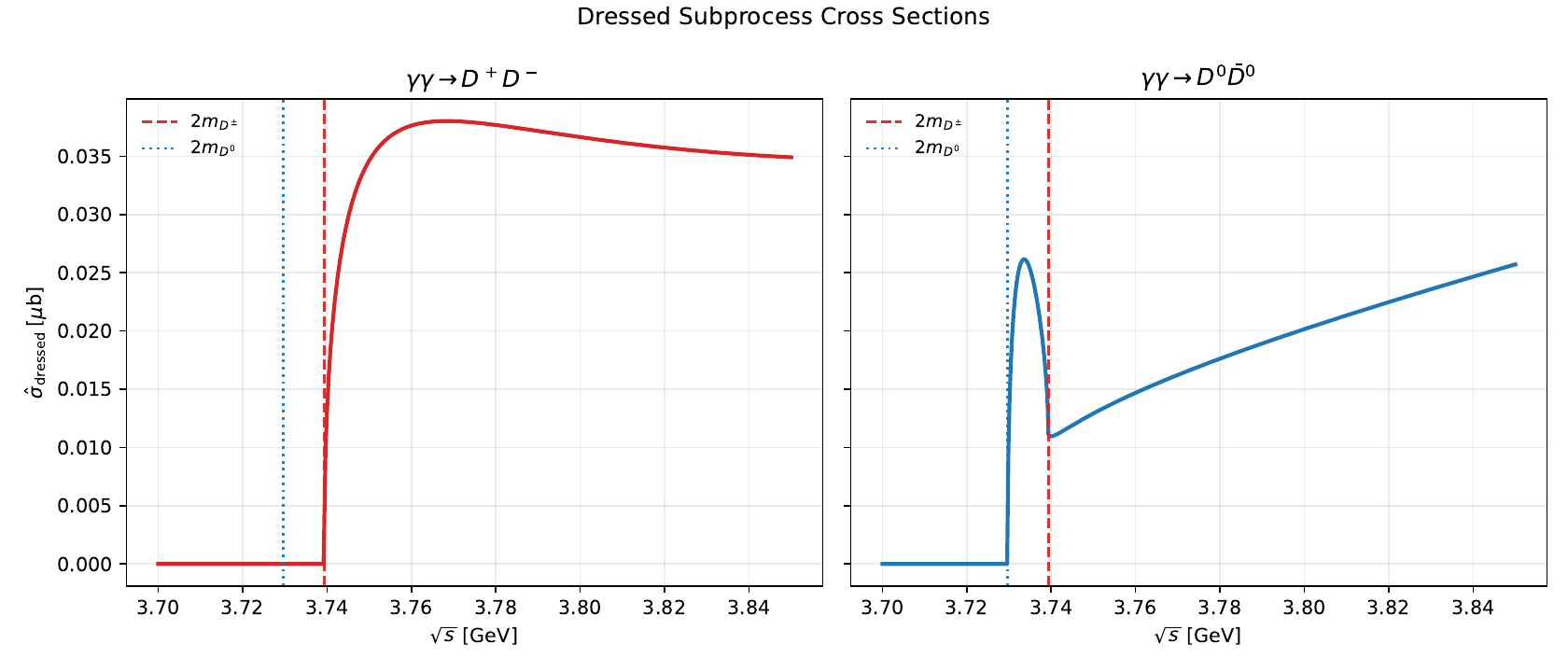}\\
\vspace{0.3cm}
%
\includegraphics[width=1\linewidth]{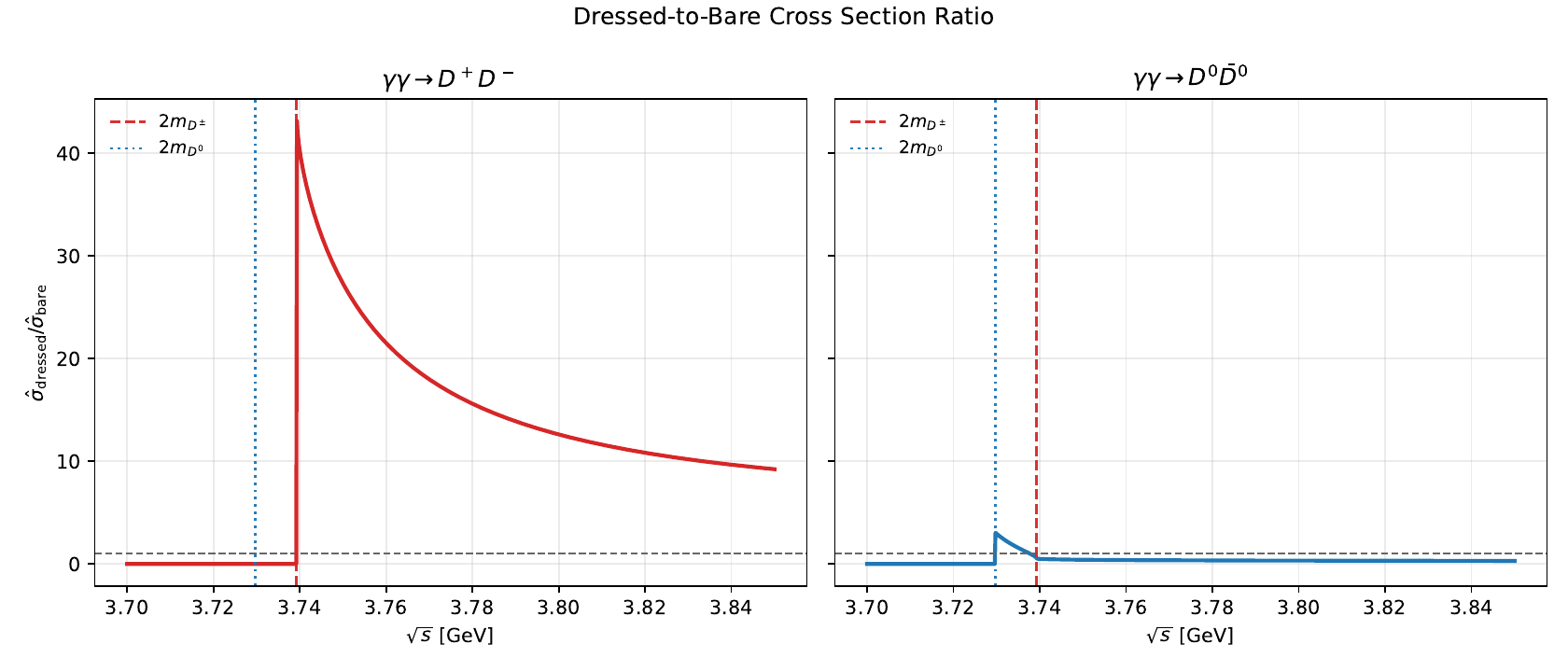}
\caption{Dressed subprocess cross sections 
$\hat{\sigma}(\gamma\gamma\to D\bar{D})$ (top). 
Ratio of dressed to bare subprocess cross sections in each channel.          
The vertical dashed lines indicate the $D^0\bar{D}^0$ and $D^+D^-$ thresholds
(bottom). The rapid charged-channel enhancement and neutral-channel suppression 
are the characteristic line-shape distortions produced by 
the adopted $X(3700)$ final-state interaction.}
\label{fig:subprocess}
\end{figure}

We first consider the elementary subprocess before convolution with the photon fluxes. Figure~\ref{fig:subprocess} shows $\hat\sigma(\gamma\gamma\to D\bar D;W)$ and the dressed-to-bare ratio. The final-state interaction enhances the charged cross section and suppresses the neutral one. The rapid variation near $W\simeq2m_{D^0}$ occurs in the same region as the enhancement of the Bethe-Salpeter amplitude discussed in Sec.~\ref{sec:fsi}.

%
%

\subsection{Comparison with Belle and BABAR}

Before embedding the signal in UPC and EIC photon fluxes, we compare the elementary calculation with the exclusive $e^+e^-\to e^+e^-D\bar D$ spectra measured by Belle and BABAR~\cite{belle_gammagamma_ddbar,babar_gammagamma_ddbar}. We extract 10-MeV bins from the vector versions of the published figures. BABAR Fig.~10 provides an efficiency-corrected candidate spectrum. After undoing the weight-rescaling constant, the reported weighted yield $76\pm17$ corresponds to $N_{\epsilon}=(285\pm64)\times10^3$ events in $384~\mathrm{fb}^{-1}$, so one plotted entry per 10-MeV bin represents approximately $0.9766~\mathrm{pb/GeV}$. Since a residual background remains in the published spectrum, the comparison provides an effective normalization constraint, not a precision cross section.

For each model prediction $t_i$ we determine the multiplicative factor $K$ over $3.74\leq W\leq3.85$~GeV from
\begin{equation}
 K=\frac{\sum_i t_i d_i/\delta_i^2}{\sum_i t_i^2/\delta_i^2},
 \qquad
 \delta K=\left(\sum_i t_i^2/\delta_i^2\right)^{-1/2},
 \label{eq:bfactory_k_fit}
\end{equation}
where $d_i$ and $\delta_i$ are the extracted BABAR values and statistical uncertainties. We obtain
\begin{equation}
 K_{\rm bare}=0.338\pm0.026,
 \qquad
 K_{\rm dressed}=0.382\pm0.029,
 \label{eq:bfactory_k_values}
\end{equation}
The corresponding values are $\chi^2/\mathrm{ndof}=11.45/10$ and $8.08/10$. Varying $Q^2_{\max}$ from $0.5$ to $2.0~\mathrm{GeV}^2$ and the upper fit boundary from $3.83$ to $3.85$~GeV gives $K_{\rm dressed}\simeq0.34$--$0.42$. Because the published BABAR spectrum contains residual background, we use this comparison only to fix the overall normalization. In the calculations below we take $K=0.382$ and use $0.34$--$0.42$ as the normalization range for absolute cross sections. The same factor multiplies the bare and dressed results and therefore cancels in their ratio.

\begin{figure}[htbp]
\includegraphics[width=\linewidth]{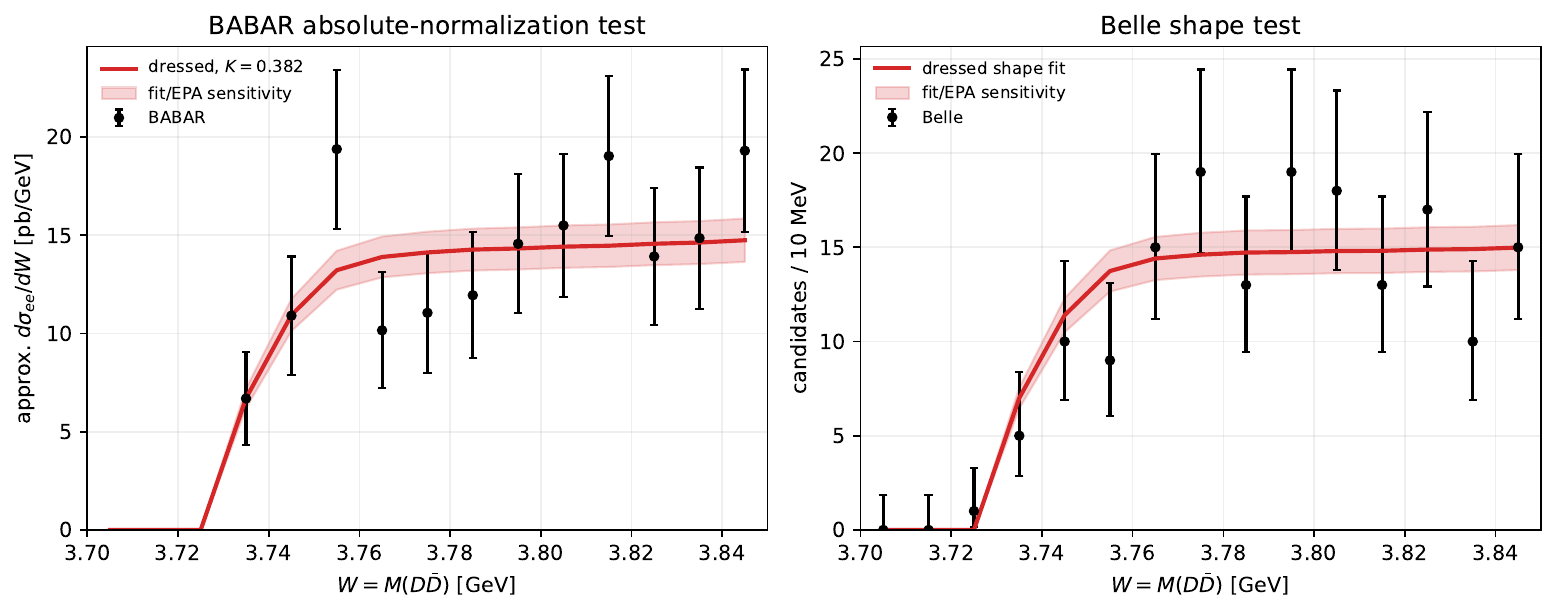}
\caption{Comparison with BABAR and Belle for $W\leq3.85$~GeV. Left: the approximate BABAR efficiency-corrected spectrum and the dressed result with $K_{\rm dressed}=0.382$. Right: the Belle event counts and the dressed shape with a fitted normalization. The red bands show the changes produced by the fit endpoint, $Q^2_{\max}$, and the normalization uncertainty; they are not confidence intervals.}
\label{fig:bfactory_normalization}
\end{figure}

Belle Fig.~2(c) gives observed event counts rather than an unfolded cross section. We therefore compare only the shape, including the reported 10\% efficiency decrease between 3.80 and 4.20~GeV and fitting one overall normalization. In Fig.~\ref{fig:bfactory_normalization}, the bands show the ranges obtained by varying the fit window and EPA inputs, together with the normalization uncertainty.

\subsection{UPC and EIC observables}

\begin{figure}[htbp]
\centering
\includegraphics[width=\linewidth]{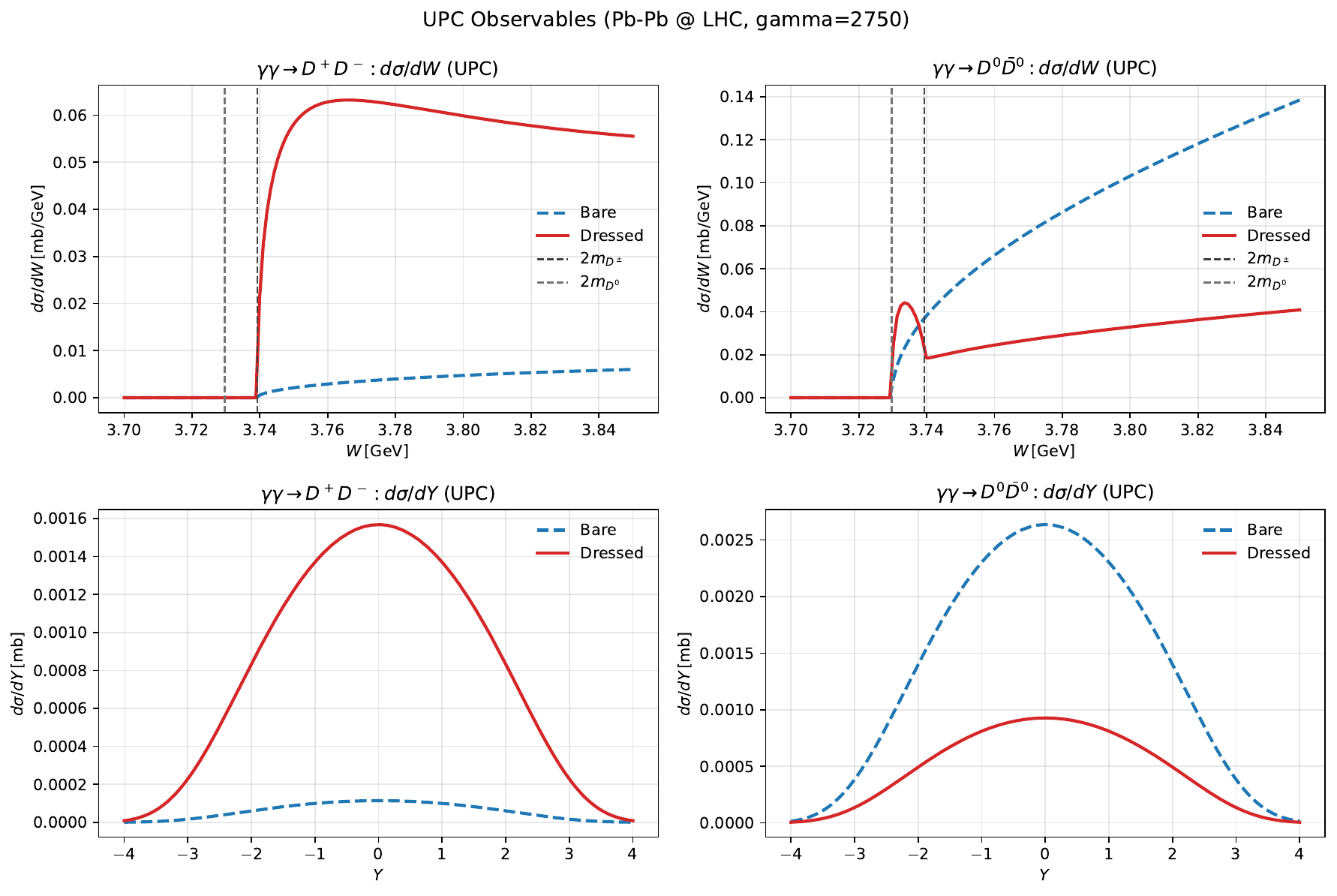} 
\caption{Invariant-mass $d\sigma/dW$ (top row) and rapidity $d\sigma/dY$ (bottom row) distributions for $D^+D^-$ (left column) and $D^0\bar{D}^0$ (right column) production in ultraperipheral Pb-Pb collisions at $\sqrt{s_{NN}}=5.5$~TeV. Dashed blue curves are bare and solid red curves are Bethe-Salpeter dressed. The corresponding BABAR-normalized absolute scale uses $K=0.382$.}
\label{fig:upc_2x2}
\end{figure}

\begin{figure}[htbp]
\centering
\includegraphics[width=\linewidth]{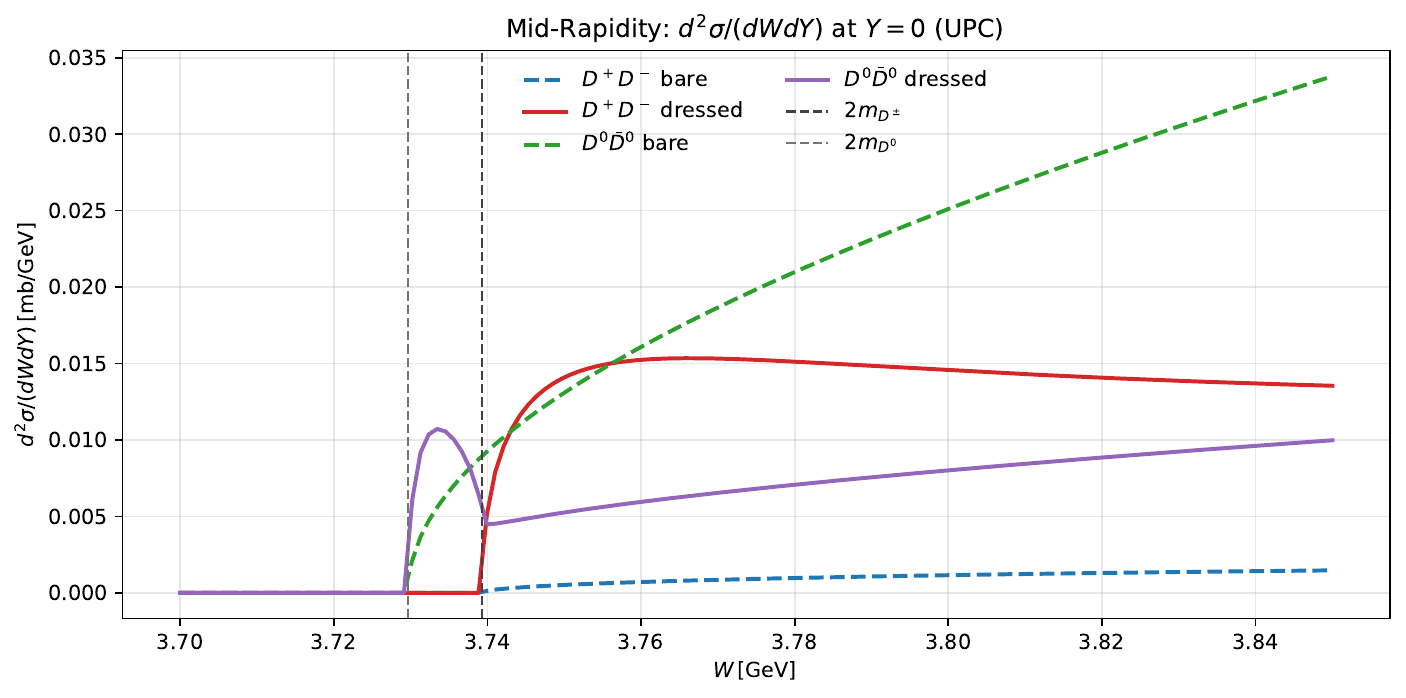}
\caption{Midrapidity ($Y=0$) differential cross section $d^2\sigma/(dW\,dY)|_{Y=0}$ for UPC Pb-Pb, comparing Born and dressed results for charged and neutral channels.}
\label{fig:upc_midrapidity}
\end{figure}

We next convolute the subprocess cross sections with the photon spectra. Figures~\ref{fig:upc_2x2} and~\ref{fig:upc_midrapidity} show the UPC invariant-mass, rapidity, and midrapidity distributions. The interval $3.70\leq W\leq3.85$~GeV contains the neutral threshold at $3.730$~GeV and the charged threshold at $3.739$~GeV, while remaining below the structure near $3.86$~GeV.

\begin{figure}[htbp]
\centering
\includegraphics[width=\linewidth]{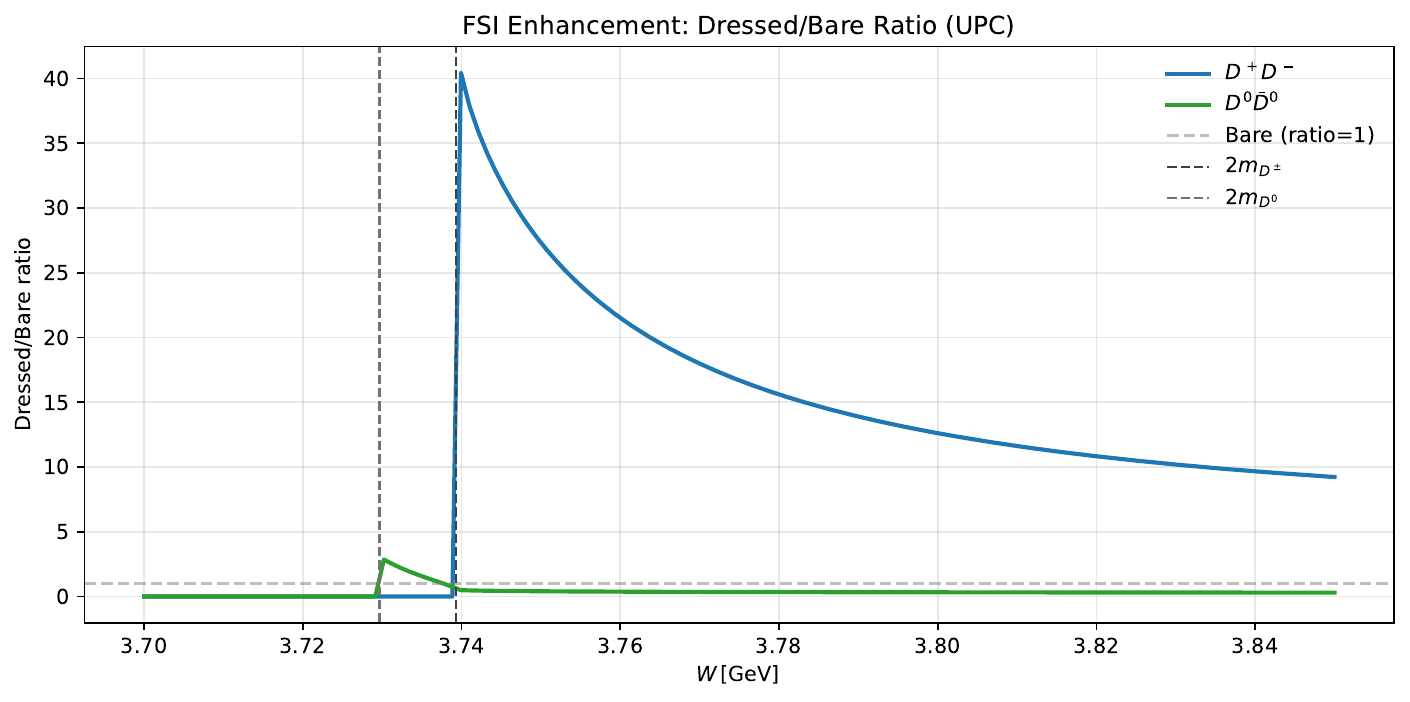}
\caption{FSI ratio $R_{\rm FSI}(W) = (d\sigma/dW)_{\rm dressed}/(d\sigma/dW)_{\rm bare}$ for $D^+D^-$ and $D^0\bar{D}^0$ in UPC Pb-Pb collisions. The common normalization cancels.}
\label{fig:upc_fsi_ratio}
\end{figure}

Figure~\ref{fig:upc_fsi_ratio} shows $R_{\rm FSI}(W)=[d\sigma/dW]_{\rm dressed}/[d\sigma/dW]_{\rm bare}$ for both channels. The common normalization $K=0.382$ cancels in this ratio. The threshold structure is therefore less sensitive to the overall normalization than the absolute cross sections.

\begin{figure}[htbp]
\centering
\includegraphics[width=\linewidth]{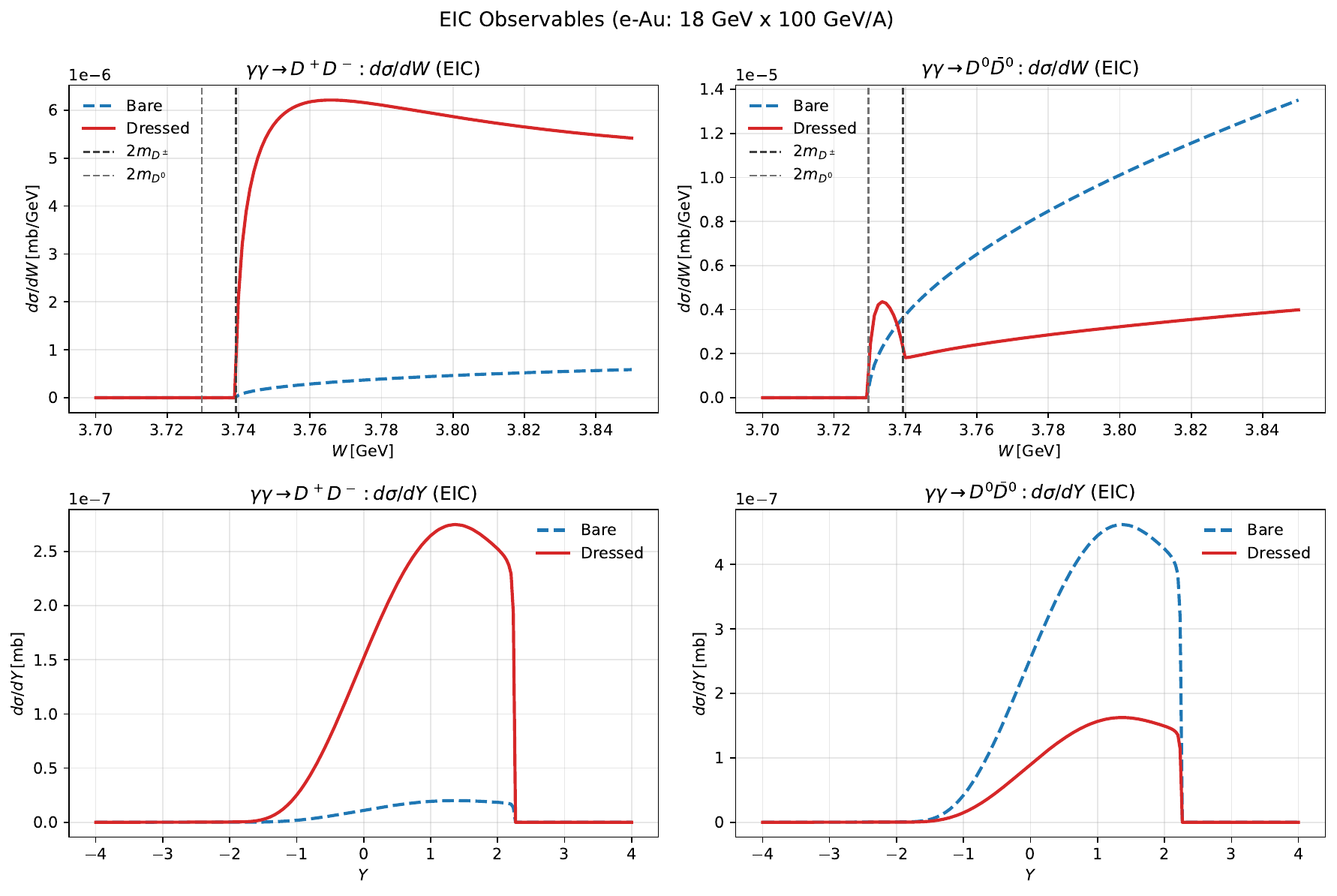}
\caption{Same as Fig.~\ref{fig:upc_2x2} but for $e$-Au collisions at the EIC ($E_e=18$~GeV, $E_A/A=100$~GeV per nucleon). The sharp $d\sigma/dY$ endpoints reflect the finite kinematic support from the $W$ window and the condition $\omega_e<E_e$ in this asymmetric-beam configuration.}
\label{fig:eic_2x2}
\end{figure}

\begin{figure}[htbp]
\centering
\includegraphics[width=\linewidth]{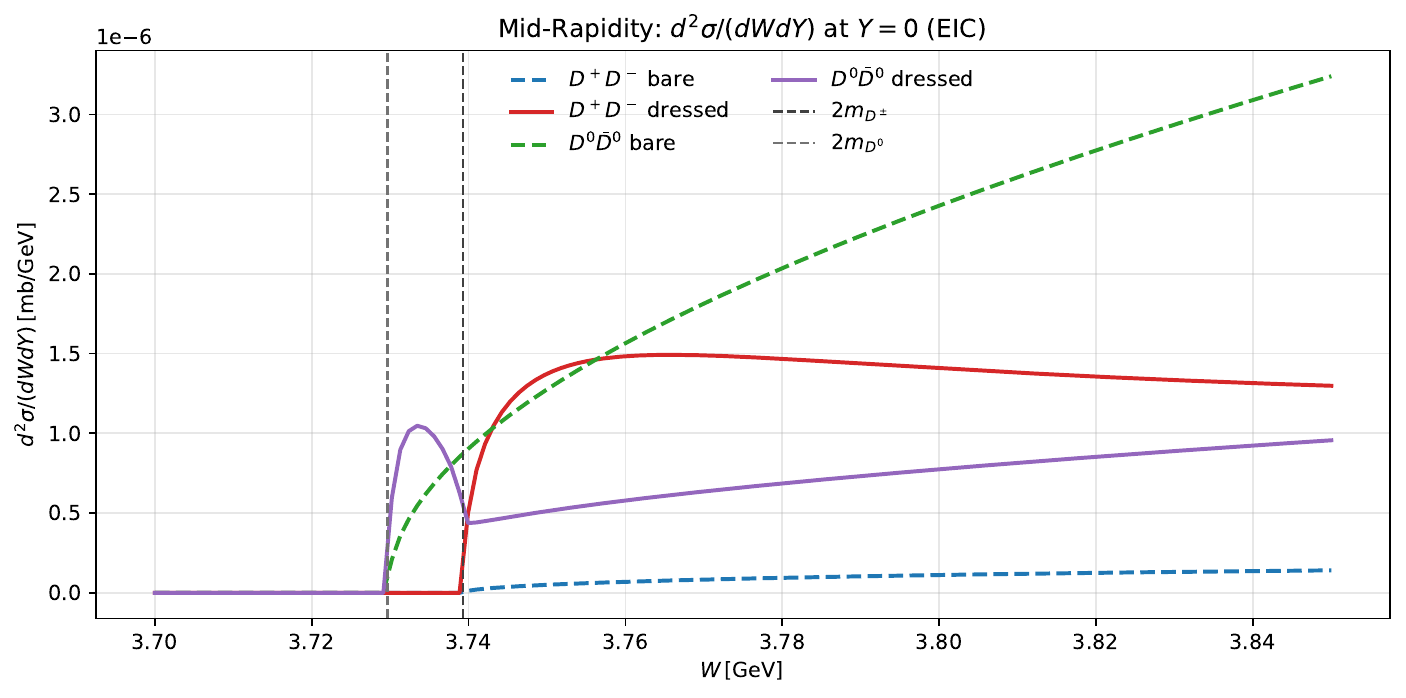}
\caption{Midrapidity differential cross section for EIC $e$-Au, comparing Born and dressed results.}
\label{fig:eic_midrapidity}
\end{figure}

The EIC spectra are shown in Figs.~\ref{fig:eic_2x2} and~\ref{fig:eic_midrapidity}. Because $E_e\ll E_A/A$, the rapidity distribution is asymmetric. In our convention it is shifted toward positive $Y$, the electron-photon side. The endpoints follow from the condition $\omega_e=(W/2)e^Y<E_e$ and from the finite range of the nuclear photon spectrum. The UPC and EIC cross sections have different normalizations, but their dressed-to-bare ratios are very similar because the same photon flux multiplies the bare and dressed subprocesses at fixed $W$.

The integrated cross sections are given in Table~\ref{tab:integrated_cross_sections}. With $K=0.382$, the charged and neutral dressed-to-bare ratios are $(13.71,0.35)$ in UPCs and $(13.74,0.35)$ at the EIC. Their agreement shows that the difference between the two channels is produced mainly by the final-state interaction.

\begin{table}[htbp]
\centering
\caption{Integrated cross sections for $D\bar{D}$ production via photon fusion in UPC (Pb-Pb) and EIC (e-Au) configurations. We use the BABAR dressed central normalization $K=0.382$ for both results.}
\label{tab:integrated_cross_sections}
\begin{tabular}{llccc}
\hline\hline
Configuration & Channel &  $\sigma_{\rm bare}$ ($\mu$b) &  $\sigma_{\rm dressed}$ ($\mu$b) & Ratio \\
\hline
\multirow{2}{*}{UPC (Pb-Pb)} & $D^+D^-$ &  $1.80$ &  $24.6$ & $13.71$ \\
& $D^0\bar{D}^0$ &  $41.3$ &  $14.6$ & $0.35$ \\
\hline
\multirow{2}{*}{EIC (e-Au)} & $D^+D^-$ &  $1.76\times 10^{-4}$ &  $2.41\times 10^{-3}$ & $13.73$ \\
& $D^0\bar{D}^0$ &  $4.05\times 10^{-3}$ &  $1.43\times 10^{-3}$ & $0.35$ \\
\hline\hline
\end{tabular}
\end{table}

\subsection{Sensitivity to model and flux inputs}

We vary the main model parameters and photon-flux inputs to identify which ones have the largest effect. These variations are not used to construct a statistical uncertainty band. We define
\begin{equation}
R_i =
\frac{\int_{3.70}^{3.85} dW\,\hat\sigma_i^{\rm dressed}(W)}
     {\int_{3.70}^{3.85} dW\,\hat\sigma_i^{\rm bare}(W)} ,
\qquad i={\rm ch,neu}.
\label{eq:subprocess_ratio_R}
\end{equation}
For the baseline parameters, $R_{\rm ch}=13.64$ and $R_{\rm neu}=0.350$. Varying the charged electromagnetic coupling over $-0.040\leq g_{\gamma,+}\leq-0.030$~GeV$^{-1}$ and the neutral electromagnetic coupling over $0.125\leq g_{\gamma,0}\leq0.159$~GeV$^{-1}$ gives $8.16\leq R_{\rm ch}\leq21.57$. The largest change comes from $g_{\gamma,0}$ because the neutral production amplitude contributes to the charged final state through the channel-changing interaction. Varying $f_D$ from $155$ to $175$~MeV gives $11.83\leq R_{\rm ch}\leq15.84$ and $0.236\leq R_{\rm neu}\leq0.572$, and shifts the threshold structure by at most about $8.9$~MeV.

The subtraction constant and scale are not varied independently. For equal-mass loops, they enter through the combination $\alpha_H+\ln(m_D^2/\mu^2)$. We therefore correlate them according to
\begin{equation}
 \alpha_H(\mu)=\alpha_H(\mu_0)+2\ln\!\left(\frac{\mu}{\mu_0}\right),
 \qquad (\alpha_H(\mu_0),\mu_0)=(-1.3,1.5~\mathrm{GeV}).
 \label{eq:alpha_mu_correlation}
\end{equation}
The correlated pairs $(\alpha_H,\mu)=(-1.586,1.3~\mathrm{GeV})$ and $(-1.050,1.7~\mathrm{GeV})$ leave the loop function and the resulting line shapes unchanged to numerical precision, providing a consistency check of the renormalization prescription rather than an additional uncertainty.

Finally, we vary the EIC photon-flux inputs. Changing $Q^2_{\max}$ from $0.5$ to $2.0$~GeV$^2$ changes the total dressed cross section by $\pm4.34\%$, while replacing the hard-sphere$\times$Yukawa form factor by a monopole increases it by $14.55\%$. The combined-channel ratio
$R_{\rm tot}=(\sigma_{\rm ch}^{\rm dressed}+\sigma_{\rm neu}^{\rm dressed})/(\sigma_{\rm ch}^{\rm bare}+\sigma_{\rm neu}^{\rm bare})$
remains approximately $0.910$. Thus the electromagnetic couplings and $f_D$ mainly control the final-state-interaction pattern, while the EPA inputs mainly change the overall normalization.

\subsection{Bound-state production estimate from Low's formula}
We separately estimate direct on-shell production, $\gamma\gamma\to X(3700)$, using Low's formula~\cite{low1960,bertulani_eic}. This pole approximation is distinct from the main dressed-continuum calculation, $\gamma\gamma\to D\bar D$, and expresses the bound-state cross section in terms of an effective two-photon width. For a scalar state of mass $M_X$,
\begin{equation}
\hat\sigma_{\gamma\gamma\to X}(W)
=8\pi^2\,\frac{\Gamma_{X\to\gamma\gamma}}{M_X}\,
\delta(W^2-M_X^2) \, .
\label{eq:low_x3700}
\end{equation}
For a dynamically generated bound state, the effective two-photon width depends on the pole residues and on how the photons couple to the meson components~\cite{branz2011}. We attach the photons to the bare $D\bar D$ production amplitudes and project them onto the bound-state. The factors $G_jg_j$ perform this projection. The resulting quantity is an effective pole residue and should not be interpreted as an elementary $X\gamma\gamma$ coupling.

Using $M_X=3.7162$~GeV, $g_{X,+-}=8.512$~GeV, and $g_{X,00}=8.604$~GeV, we estimate the effective electromagnetic pole residue as
\begin{equation}
g_{\gamma\gamma}^{\rm eff}
=\sum_{j=+-,00} g_{X,j}\,G_j(\hat{s}_X)\,\mathcal M_j^{\rm bare}(\hat{s}_X) \, .
\label{eq:ggamma_eff_residue}
\end{equation}
With the present production operator this gives $g_{\gamma\gamma}^{\rm eff}\simeq -0.0413\,i$~GeV and
\begin{equation}
\Gamma_{X\to\gamma\gamma}^{\rm raw}
=\frac{|g_{\gamma\gamma}^{\rm eff}|^2}{16\pi M_X}
=9.14~{\rm keV}.
\label{eq:x3700_gamma_gamma_width_raw}
\end{equation}
The decomposition is strongly channel dependent:
\begin{equation}
\Gamma_{\rm ch}=0.270~{\rm keV},\qquad
\Gamma_{\rm neu}=6.264~{\rm keV},\qquad
\Gamma_{\rm int}=2.602~{\rm keV}.
\end{equation}
This decomposition shows the strong channel dependence of the Low-formula estimate.

Folding Eq.~\eqref{eq:low_x3700} with the same EIC $e$-Au photon fluxes gives the cross sections in Table~\ref{tab:low_x3700}. The result is proportional to the effective two-photon width. With the BABAR normalization, we obtain $\Gamma_{\gamma\gamma}=3.49$~keV and $\sigma_{\rm EIC}^{\rm Low}=6.60$~nb.

\begin{table}[htbp]
\centering
\small
\caption{EIC $e$-Au cross sections for direct $X(3700)$ bound-state production obtained from Low's formula and the effective two-photon width.}
\label{tab:low_x3700}
\begin{tabular}{lcc}
\hline\hline
Scenario & $\Gamma_{X\to\gamma\gamma}$ (keV) & $\sigma_{\rm EIC}^{\rm Low}$ (nb) \\
\hline
Without BABAR normalization ($K=1$) & 9.14 & 17.28 \\
BABAR normalization ($K=0.382$) & 3.49 & 6.60 \\
Normalization range & 3.11--3.84 & 5.88--7.26 \\
\hline\hline
\end{tabular}
\end{table}
\subsection{Event yields and experimental considerations}

To estimate the number of produced events, we use the dressed cross sections in the same mass and rapidity window. These estimates do not include charm-hadron branching fractions, detector acceptance, reconstruction efficiencies, analysis cuts, or backgrounds. The number of produced events is

\begin{equation}
    N = \mathcal{L}_{\rm int} \cdot \sigma_{\rm dressed},
    \label{eq:n_vis}
\end{equation}
Here $\mathcal{L}_{\rm int}$ is the integrated luminosity. Table~\ref{tab:phenomenology_event_yields} reports $N$ from Eq.~\eqref{eq:n_vis}. For the UPC estimates we use $1.5$~nb$^{-1}$, the total Pb--Pb luminosity delivered to ALICE during Run~2, and $3.9$~nb$^{-1}$, the cumulative luminosity delivered in the 2023 and 2024 Pb--Pb runs ($2.0+1.9$~nb$^{-1}$)~\cite{bruce_ipac2025}. These luminosities are used only as reference values; the UPC cross sections are evaluated at $\sqrt{s_{NN}}=5.5$~TeV as specified in this work. For the EIC we use $10$ and $100$~fb$^{-1}$, motivated by Ref.~\cite{eic_yellow_report}.

\begin{table}[htbp]
\centering
\small
\caption{Produced event counts obtained from $N = \mathcal{L}_{\rm int}\,\sigma_{\rm dressed}$ for Run~2 and Run~3 Pb--Pb luminosity benchmarks at the LHC and projected EIC luminosities. These numbers are before branching fractions, acceptance, reconstruction efficiencies, analysis cuts, and backgrounds. Our results use the BABAR dressed central normalization $K=0.382$.}
\label{tab:phenomenology_event_yields}
\begin{tabular}{llccc}
\hline\hline
Configuration & Channel & $\mathcal{L}_{\rm int}$ &  $\sigma_{\rm dressed}$ ($\mu$b) &  $N$ \\
\hline
UPC Run~2 & $D^+D^-$ & $1.5\,\mathrm{nb}^{-1}$ &  $24.6$ &  $3.69\times 10^{4}$ \\
UPC Run~3 (2023+2024) & $D^+D^-$ & $3.9\,\mathrm{nb}^{-1}$ &  $24.6$ &  $9.59\times 10^{4}$ \\
UPC Run~2 & $D^0\bar{D}^0$ & $1.5\,\mathrm{nb}^{-1}$ &  $14.6$ &  $2.19\times 10^{4}$ \\
UPC Run~3 (2023+2024) & $D^0\bar{D}^0$ & $3.9\,\mathrm{nb}^{-1}$ &  $14.6$ &  $5.69\times 10^{4}$ \\
EIC & $D^+D^-$ & $10\,\mathrm{fb}^{-1}$ &  $2.41\times 10^{-3}$ &  $2.41\times 10^{7}$ \\
EIC & $D^+D^-$ & $100\,\mathrm{fb}^{-1}$ &  $2.41\times 10^{-3}$ &  $2.41\times 10^{8}$ \\
EIC & $D^0\bar{D}^0$ & $10\,\mathrm{fb}^{-1}$ &  $1.43\times 10^{-3}$ &  $1.43\times 10^{7}$ \\
EIC & $D^0\bar{D}^0$ & $100\,\mathrm{fb}^{-1}$ &  $1.43\times 10^{-3}$ &  $1.43\times 10^{8}$ \\
\hline\hline
\end{tabular}
\end{table}

The charged and neutral channels offer different reconstruction possibilities~\cite{PDG2024,pedro_decays}. The decay $D^+\to K^-\pi^+\pi^+$ gives an all-charged final state, while the neutral channel can be reconstructed, for example, through $D^0\to K^-\pi^+\pi^-\pi^+$. A quantitative assessment of branching fractions, efficiencies, and backgrounds requires a detector-level simulation and is beyond the scope of the present work. The exclusive two-photon topology and the restricted near-threshold mass interval may nevertheless assist the experimental selection.
 \section{Conclusions}
\label{sec:conclusions}

We have studied near-threshold $D\bar D$ production through two-photon fusion in ultraperipheral Pb--Pb collisions at the LHC and in $e$-Au collisions at the EIC. We calculated the charged and neutral tree-level production amplitudes, projected them onto the $S$ wave, and included the final-state interaction by solving the two-channel Bethe-Salpeter equation in the $\{D^+D^-,D^0\bar D^0\}$ basis. The resulting subprocess cross sections were then convoluted with the corresponding equivalent-photon spectra.

With the interaction and parameters used here, the analytic continuation of the $T$-matrix gives a bound state at $M_X=3.7162$~GeV, below both open-charm thresholds. Its effect on the continuum is different in the two channels: the charged cross section is enhanced, while the neutral cross section is suppressed. This behavior remains visible after convolution with the UPC and EIC photon fluxes. The integrated dressed-to-bare ratios are $(13.71,0.35)$ in UPCs and $(13.74,0.35)$ at the EIC for the charged and neutral channels, respectively.

The absolute normalization is constrained by comparison with the BABAR spectrum, for which we obtain $K_{\rm dressed}=0.382\pm0.029$. The central value gives dressed cross sections of $(24.6,14.6)~\mu\mathrm b$ in UPCs and $(2.41,1.43)\times10^{-3}~\mu\mathrm b$ at the EIC for $(D^+D^-,D^0\bar D^0)$; these remain normalization-dependent estimates because the published spectrum contains residual background. The separate Low-formula calculation of direct on-shell production gives $\sigma_{\rm EIC}^{\rm Low}=6.60$~nb for the same normalization.

The electromagnetic $\gamma DD^*$ couplings and $f_D$ produce the largest changes in the channel ratios, whereas the EPA inputs mainly affect the overall normalization. The bound-state interpretation remains dependent on the adopted interaction kernel and the two-channel approximation; additional coupled channels and higher partial waves are beyond the present calculation.

Charged-to-neutral ratios are directly accessible observables, while dressed-to-bare ratios quantify the role of the final-state interaction within the model. Measurements in UPCs and at the EIC would provide information complementary to the existing Belle and BABAR data. Future refinements may include additional coupled channels and higher partial waves.

\begin{acknowledgments}
This work was partially financed by the Brazilian funding agencies CNPq, CAPES, FAPESP and INCT-FNA (process number 464898/2014-5). The authors gratefully acknowledge the support from the Funda\c{c}\~ao de Amparo \`a Pesquisa do Estado de S\~ao Paulo (FAPESP), process number 2024/19103-9.
\end{acknowledgments}

\appendix
\section{Projected amplitudes for $\gamma\gamma\to D\bar{D}$}
\label{app:projection}

This appendix gives the analytic $J=0$ projections of the charged and neutral production amplitudes used in the channel-resolved cross sections of Eq.~\eqref{eq:sigmahat}.

\subsection*{Kinematics and helicity amplitudes}

In the $\gamma\gamma$ center-of-mass frame, the Mandelstam variables are
\begin{equation}
	\hat{t} = m_D^2 - \frac{\hat{s}}{2}(1-\beta_D\cos\theta) \, ,\qquad
	\hat{u} = m_D^2 - \frac{\hat{s}}{2}(1+\beta_D\cos\theta) \, ,
\end{equation}
where $\theta$ is the scattering angle and $\beta_D=\sqrt{1-4m_D^2/\hat{s}}$. For either physical channel, the $S$-wave projection is
\begin{equation}
	\mathcal{M}_0(\hat{s}) = \frac{1}{2}\int_{-1}^{1} d(\cos\theta)\,\mathcal{M}(\hat{s},\cos\theta) \, .
	\label{eq:swave_def}
\end{equation}
We choose the four-momenta as
\begin{align}
	k^\mu &= \frac{\sqrt{\hat{s}}}{2}(1,0,0,1) \, , &
	k'^\mu &= \frac{\sqrt{\hat{s}}}{2}(1,0,0,-1) \, , \\
	p^\mu &= \left(\frac{\sqrt{\hat{s}}}{2},\,p\sin\theta,\,0,\,p\cos\theta\right) \, , &
	p'^\mu &= \left(\frac{\sqrt{\hat{s}}}{2},\,-p\sin\theta,\,0,\,-p\cos\theta\right) \, ,
\end{align}
with $p=\tfrac{1}{2}\sqrt{\hat{s}-4m_D^2}$. The $J=0$ projection receives contributions from the two-photon state with total helicity zero. For counterpropagating photons, this corresponds to equal photon helicities,
\begin{equation}
	\mathcal{M}_{\lambda\lambda}(\hat{s},\theta)
	\equiv \mathcal{M}\bigl[\gamma(k,\lambda)\gamma(k',\lambda)\to D\bar{D}\bigr] \, ,
	\qquad \lambda=\pm 1 \, .
\end{equation}
For $\lambda=+1$, we use
\begin{equation}
	\varepsilon^\mu(k,+)=\frac{1}{\sqrt{2}}(0,-1,-i,0) \, ,
	\qquad
	\varepsilon^\mu(k',+)=\frac{1}{\sqrt{2}}(0,1,-i,0) \, ,
\end{equation}
which gives
\begin{equation}
	\varepsilon(k,+)\!\cdot\!\varepsilon(k',+)=-1 \, ,
	\qquad
	\varepsilon(k,+)\!\cdot\!p=-\frac{p\sin\theta}{\sqrt{2}} \, ,
	\qquad
	\varepsilon(k',+)\!\cdot\!p=\frac{p\sin\theta}{\sqrt{2}} \, .
\end{equation}
Parity gives $\mathcal{M}_{++}=\mathcal{M}_{--}$. The opposite-helicity amplitudes have $|J_z|=2$ and begin at $J=2$, so they do not contribute to the $S$ wave.

The angular integral of the charged-$D$ propagator is
\begin{equation}
	\frac{1}{2}\int_{-1}^{1}\frac{d(\cos\theta)}{\hat{t}-m_D^2}
	= \frac{1}{2}\int_{-1}^{1}\frac{-2\,d(\cos\theta)}{\hat{s}(1-\beta_D\cos\theta)}
	= \frac{-2\,\mathrm{arctanh}(\beta_D)}{\hat{s}\,\beta_D} \, .
	\label{eq:D_propagator_integral}
\end{equation}
For $D^*$ exchange, $\hat{t}-m_{D^*}^2=m_D^2-m_{D^*}^2-\frac{\hat{s}}{2}(1-\beta_D\cos\theta)$, and the corresponding integral is
\begin{equation}
	\frac{1}{2}\int_{-1}^{1}\frac{d(\cos\theta)}{\hat{t}-m_{D^*}^2}
	= -\frac{2\,\mathrm{arccoth}(z^*)}{\sqrt{\hat{s}(\hat{s}-4m_D^2)}} \, ,
	\label{eq:Dstar_propagator_integral}
\end{equation}
where $z^*=(2m_{D^*}^2-2m_D^2+\hat{s})/\sqrt{\hat{s}(\hat{s}-4m_D^2)}$. The $u$-channel integral has the same form after $\cos\theta\to-\cos\theta$.

\subsection*{Charged and neutral projections}

For the charged channel, the five tree-level diagrams of Sec.~\ref{sec:production} give the equal-helicity amplitude
\begin{equation}
	\mathcal{M}_{++,+-}(\hat{s},\theta) = -2ie^2m_{D^\pm}^2
	\left(\frac{1}{\hat{t}-m_{D^\pm}^2} + \frac{1}{\hat{u}-m_{D^\pm}^2}\right)
	+ 8ig_{\gamma,+}^2\hat{s}
	+ 4ig_{\gamma,+}^2m_{D^{*\pm}}^2\hat{s}
	\left(\frac{1}{\hat{t}-m_{D^{*\pm}}^2} + \frac{1}{\hat{u}-m_{D^{*\pm}}^2}\right) \, .
	\label{eq:helicity_zero_charged}
\end{equation}
The scalar-QED contact term cancels the angle-independent part of the charged-$D$ exchange contribution. The constant term $8ig_{\gamma,+}^2\hat{s}$ comes from the longitudinal part of the $D^*$ propagator and is unchanged by the angular integration.

Using Eqs.~\eqref{eq:swave_def}, \eqref{eq:D_propagator_integral}, and \eqref{eq:Dstar_propagator_integral}, we obtain
\begin{equation}
	\mathcal{M}_{0,+-}(\hat{s}) = 4i\left[
	\frac{2e^2 m_{D^\pm}^2\,\mathrm{arctanh}(\beta_{D^\pm})
		  - 4g_{\gamma,+}^2 m_{D^{*\pm}}^2\hat{s}\,\mathrm{arccoth}(z^*)}
		 {\sqrt{\hat{s}(\hat{s}-4m_{D^\pm}^2)}}
	+ 2g_{\gamma,+}^2\hat{s}
	\right] \, ,
	\label{eq:M0ch_app}
\end{equation}
Here $z^*=(2m_{D^{*\pm}}^2-2m_{D^\pm}^2+\hat{s})/\sqrt{\hat{s}(\hat{s}-4m_{D^\pm}^2)}$ and $\mathrm{arccoth}(z)=\frac{1}{2}\ln[(z+1)/(z-1)]$, with the appropriate analytic continuation. The last term comes from the longitudinal part of the $D^*$ propagator.

For the neutral channel, only $D^{*0}$ exchange contributes, giving
\begin{equation}
	\mathcal{M}_{++,00}(\hat{s},\theta) = 8ig_{\gamma,0}^2\hat{s}
	+ 4ig_{\gamma,0}^2m_{D^{*0}}^2\hat{s}
	\left(\frac{1}{\hat{t}-m_{D^{*0}}^2} + \frac{1}{\hat{u}-m_{D^{*0}}^2}\right) \, .
	\label{eq:helicity_zero_neutral}
\end{equation}
Its $S$-wave projection is
\begin{equation}
	\mathcal{M}_{0,00}(\hat{s}) = -4ig_{\gamma,0}^2\hat{s}\left[
	\frac{4m_{D^{*0}}^2\,\mathrm{arccoth}(z_0)}
		 {\sqrt{\hat{s}(\hat{s}-4m_{D^0}^2)}}
	- 2
	\right] \, ,
	\label{eq:M0n_app}
\end{equation}
where $z_0=(2m_{D^{*0}}^2-2m_{D^0}^2+\hat{s})/\sqrt{\hat{s}(\hat{s}-4m_{D^0}^2)}$.

With the explicit overall factor $i$ in our convention, Eqs.~\eqref{eq:M0ch_app} and~\eqref{eq:M0n_app} are purely imaginary on the physical real-$\hat{s}$ axis. The combinations containing $\mathrm{arctanh}(\beta_D)$ and $\mathrm{arccoth}(z)$ have finite threshold limits. These channel-resolved projections are used in the bare cross sections and as input to the Bethe-Salpeter dressing in Eq.~\eqref{eq:dressed_amplitude}.
 \section{Equivalent-photon calculation}
\label{app:epa}

This appendix gives the kinematic limits and nuclear-flux checks used in the UPC and EIC calculations.

\subsection*{Kinematics and photon spectra}

For $e$-Au collisions, the invariant mass $W=\sqrt{4\omega_e\omega_A}$ and rapidity $Y=\frac{1}{2}\ln(\omega_e/\omega_A)$ determine the photon energies through
\begin{equation}
	\omega_e = \frac{W}{2}\,e^{+Y} \, , \qquad \omega_A = \frac{W}{2}\,e^{-Y} \, .
	\label{eq:WY_to_omega}
\end{equation}
We use $3.70\leq W\leq3.85$~GeV and $-4\leq Y\leq4$. The upper mass limit keeps the calculation below the additional structure near $3.86$~GeV, which is not included in the present $S$-wave model. Points with $\omega_e\leq0$ or $\omega_e\geq E_e$ do not contribute, and the nuclear form factor suppresses the flux at large $\omega_A$.

In Eq.~\eqref{eq:flux_electron}, the lower virtuality limit is fixed by kinematics, $Q^2_{\min}=m_e^2\omega_e^2/[E_e(E_e-\omega_e)]$. For $\omega_e\simeq W/2\simeq1.87$~GeV and $E_e=18$~GeV, this gives $Q^2_{\min}\simeq10^{-5}$~GeV$^2$. We take $Q^2_{\max}=1.0$~GeV$^2$ to select quasireal photons.

For the EIC nuclear spectrum, we integrate the coherent flux over the complete transverse plane using the momentum-space expression~\cite{krauss1997,bertulani_eic,klein_steinberg_upc}
\begin{equation}
	f_{\gamma/A}(\omega_A)=\frac{Z^2\alpha}{\pi\omega_A}
	\int_0^\infty dk_T^2\,
	\frac{k_T^2}{\left[k_T^2+(\omega_A/\gamma_L)^2\right]^2}
	\left|F_A\!\left(\sqrt{k_T^2+(\omega_A/\gamma_L)^2}\right)\right|^2,
	\label{eq:eic_all_b_flux}
\end{equation}
This is the all-impact-parameter form of Eq.~\eqref{eq:flux_nucleus}. The elastic form factor suppresses large transverse momenta and imposes nuclear coherence. No hadronic-overlap condition is needed in $e$-Au collisions. For Pb--Pb collisions, we impose $b_{\min}=2R_A$ to exclude nuclear overlap.

\subsection*{Nuclear form-factor dependence}

The main calculation uses the hard-sphere$\times$Yukawa form factor. We compare it with point-like, monopole, and Woods-Saxon profiles. For the monopole form, we use $\Lambda=91$~MeV for Au and $88$~MeV for Pb. The cumulative Pb spectrum shows that $86.39\%$ of the photons lie below $2$~GeV and $96.85\%$ below $10$~GeV in the default model. The finite-size models differ only at the per-mille level in this energy range, explaining their small effect on the near-threshold observables.

\begin{figure}[htbp]
\centering
\includegraphics[width=0.85\linewidth]{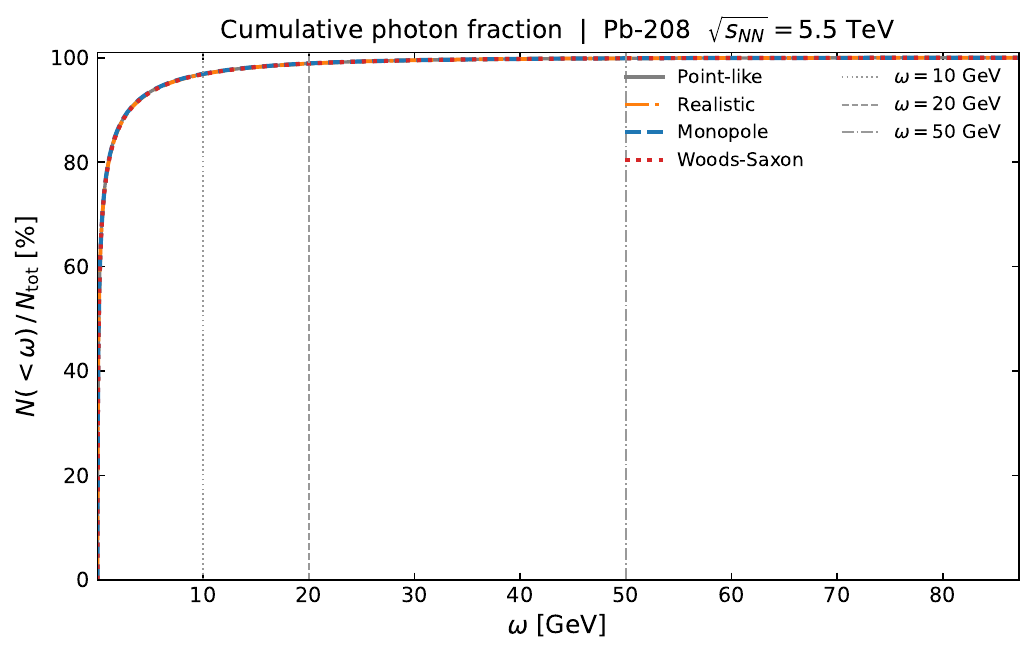}
\caption{Cumulative Pb photon fraction $N(<\omega)/N_{\rm tot}$ at $\sqrt{s_{NN}}=5.5$~TeV for the point-like, monopole, hard-sphere$\times$Yukawa, and Woods-Saxon form factors. The finite-size results are nearly identical in the energy range relevant for threshold $D\bar D$ production.}
\label{fig:photon_cumulative_fraction}
\end{figure}

\begin{table}[htbp]
\centering
\caption{Cumulative Pb photon fraction $N(<\omega)/N_{\rm tot}$ (in \%) at $\sqrt{s_{NN}}=5.5$~TeV for the different nuclear form factors.}
\label{tab:photon_cumulative_percentages}
\begin{tabular}{ccccc}
\hline\hline
$\omega$ [GeV] & Point-like & Monopole & Realistic & Woods-Saxon \\
\hline
1.0 & 79.49 & 79.50 & 79.27 & 79.27 \\
2.0 & 86.56 & 86.58 & 86.39 & 86.39 \\
5.0 & 93.47 & 93.49 & 93.38 & 93.38 \\
10.0 & 96.90 & 96.91 & 96.85 & 96.85 \\
20.0 & 98.91 & 98.91 & 98.89 & 98.89 \\
30.0 & 99.53 & 99.53 & 99.52 & 99.52 \\
50.0 & 99.89 & 99.89 & 99.89 & 99.89 \\
75.0 & 99.99 & 99.99 & 99.99 & 99.99 \\
\hline
$N_{\mathrm{tot}}$ & 9.33e+02 & 9.32e+02 & 9.19e+02 & 9.20e+02 \\
$\omega_{\mathrm{peak}}$ [GeV] & 0.010 & 0.010 & 0.010 & 0.010 \\
\hline\hline
\end{tabular}
\end{table}

\end{document}